\documentclass[11pt]{article}
\pdfoutput=1
\usepackage{jcappub,natbib}
\input{colordvi.tex}

\def\ga{\mathrel{\raise.3ex\hbox{$>$\kern-.75em\lower1ex\hbox{$\sim$}}}}
\def\la{\mathrel{\raise.3ex\hbox{$<$\kern-.75em\lower1ex\hbox{$\sim$}}}}

\def\betap{\tilde\beta}

\def\Msun{M_\odot}
\def\calO{{\cal O}(1)}

\title{Gravitational waves at interferometer scales and primordial black holes in axion inflation}

\author[a]{Juan Garc\'ia-Bellido,}
\author[b]{Marco Peloso,}
\author[b]{Caner Unal}

\affiliation[a]{Instituto de F\'isica Te\'orica UAM-CSIC, Universidad Auton\'oma de Madrid, Cantoblanco,  Madrid, 28049 (Spain)}
\affiliation[b]{School of Physics and Astronomy, and Minnesota Institute for Astrophysics, University of Minnesota, Minneapolis, 55455 (USA)}

\abstract{
We study the prospects of detection at terrestrial and space interferometers, as well as at pulsar timing array experiments, of a stochastic gravitational wave background which can be produced in models of axion inflation. This potential signal, and the development of these experiments, open a new window on inflation on scales much smaller than those currently probed with Cosmic Microwave Background and Large Scale Structure measurements. The sourced signal generated in axion inflation is an ideal candidate for such searches, since it naturally grows at small scales, and it has specific properties (chirality and non-gaussianity) that can distinguish it from an astrophysical background. We study under which conditions such a signal can be produced at an observable level, without the simultaneous overproduction of scalar perturbations in excess of what is allowed by the primordial black hole limits. We also explore the possibility that scalar perturbations generated in a modified version of this model may provide a distribution of primordial black holes compatible with the current bounds, that can act as a seeds of the present black holes in the universe.  
}

\begin{document}

\begin{flushright}  IFT--UAM/CSIC--16--100, UMN--TH--3607/16  \end{flushright}

\maketitle
\flushbottom

\section{Introduction}%
\label{sec:intro}

Cosmic Microwave Background (CMB) and Large Scale Structure (LSS) observations strongly support the paradigm of cosmological inflation~\cite{Ade:2015lrj,Cuesta:2015mqa}. However, they allow us to directly probe only a small fraction of the inflationary evolution. CMB and LSS probe the range of wave numbers $10^{-4} \, {\rm Mpc}^{-1} \la k \la 0.1  \, {\rm Mpc}^{-1}$, corresponding to about $7$ e-folds of inflation. CMB $y-$ and $\mu-$distortions may allow us to probe smaller scales,  extending the above range to $\sim 10^4\, {\rm Mpc}^{-1}$. Even this extended range would cover only $\sim 18$ out of the $50-60$ e-folds of inflation that produce perturbations within our horizon. This leaves the remaining $\sim 30-40$ e-folds largely unexplored,  apart from the bounds and the potential signatures associated with primordial black holes (PBH), that arise if the scalar perturbations have a sufficiently high amplitude at those scales~\cite{GarciaBellido:1996qt,Clesse:2015wea}. 

The recent gravitational waves (GW) observations at the Advanced LIGO detector (AdvLIGO)  \cite{Abbott:2016blz,Abbott:2016nmj} have opened a new observational window on general relativity, astrophysics, and cosmology. Concerning cosmology, GW measurements from a network of terrestrial (such as Advanced LIGO, Advanced Virgo, KAGRA, and LIGO-India), from Pulsar Timing Array experiments, and from  space interferometers (such as LISA) will provide invaluable information on galactic and stellar evolution. They will also give us the unique opportunity to probe specific models of inflation, or specific mechanisms that could have been acting during inflation, at much smaller scales than those probed by CMB acoustic peaks, LSS, and CMB distortions. 

We explore here the possibility that an enhanced GW background is produced directly during inflation, in the frequency ranges probed by terrestrial or spatial GW interferometers. For instance, AdvLIGO is mostly sensitive to GW in the frequency range $\left( 10-200 \right)$ Hz \cite{TheLIGOScientific:2016wyq}, corresponding to  wavenumbers $k \sim  \left( 10^{16} - 10^{17} \right) {\rm Mpc}^{-1}$.  LISA is instead mostly sensitive to  GW in the frequency range $\left( 10^{-4} - 10^{-1} \right)$ Hz  \cite{Caprini:2015zlo,Bartolo:2016ami} corresponding to  wavenumbers $k \sim  \left( 10^{11} - 10^{14} \right) {\rm Mpc}^{-1}$. Finally, PTA experiments are mostly sensitive to  GW in the frequency range $\left( 10^{-9} - 10^{-7} \right)$ Hz \cite{Arzoumanian:2015liz,Lentati:2015qwp,Shannon:2015ect}.~\footnote{The PTA discussion was absent in the first version of this work, and it has been added following a suggestion from an anonymous referee. MP acknowledges an e-mail exchange with Kin-Wang Ng on this subject.} $\!\!\!^{,}\!\!$~\footnote{Constrains on inflationary models from PTA experiments can be found for instance in \cite{Zhao:2013bba,Liu:2015psa}.}

A different possibility is that density perturbations produced during inflation collapse to form PBH, which then evolve to the present universe, and ultimately give rise to BH-BH binary mergers, such as those observed by the AdvLIGO detectors. This is a different mechanism of GW production, which is sensitive to modes of different scales.  AdvLIGO is sensitive to collisions between black hole binaries up to a few tens of solar masses, since the frequency of the innermost stable circular orbit $f_{\rm ISCO} = 4.4\, {\rm kHz}\, \Msun/M$ has to be above the seismic noise to be detectable by AdvLIGO~\cite{Abbott:2016bqf}. As we discuss in Section \ref{subsec:collision}, these black holes arise from merging and accretion from initial PBH seeds of masses between a few thousandth and a few hundredth of solar masses, corresponding to wavenumbers in the  $\sim \left( 10^{7} - 10^{8} \right) {\rm Mpc}^{-1}$ range.  LISA is instead sensitive to black hole binaries from $10^5$ to roughly $10^8$ solar masses \cite{Berti:2005ys,Berti:2009kk}. As we discuss in Section \ref{subsec:collision}, these black holes arise from merging and accretion from initial PBH seeds of masses in the $\sim 1-10^3 \, M_\odot$ range, corresponding to wavenumbers in the  $\sim \left( 10^{5} - 10^{7} \right) {\rm Mpc}^{-1}$ range.   PTA experiments are mostly sensitive to black hole masses from roughly $10^8$ to  roughly $10^{10}$ solar masses \cite{Moore:2014eua}, which, assuming a comparable merging to the PBH probed by LISA, corresponds to PBH seeds of masses in the $\sim 10^3-10^5 \, M_\odot$ range, and to wavenumbers in the  $\sim \left( 10^{4} - 10^{5} \right) {\rm Mpc}^{-1}$ range.

\begin{center}
\begin{table}[t]
\centering
\begin{tabular}{|l|c|c|}  \hline       
 & $k \;\; \left[ {\rm Mpc}^{-1} \right] $ & $ N_{\rm estim.} $ \\ \hline 
CMB / LSS & $10^{-4} - 10^{-1}$ & $56-63$ \\ \hline
$y-$ \& $\mu-$distortions & $10^{-1} - 10^4 $ & $45-56$ \\ \hline
$P_\zeta \,\rightarrow\, {\rm PBH} \, \rightarrow \, {\rm GW}$ @ PTA  & $10^4 - 10^5 $ & $41-44$ \\ \hline
$P_\zeta \,\rightarrow\, {\rm PBH} \, \rightarrow \, {\rm GW}$ @ LISA  & $10^5 - 10^7 $ & $38-41$ \\ \hline
$P_\zeta \,\rightarrow\, {\rm PBH} \, \rightarrow \, {\rm GW}$ @ AdvLIGO  & $ 10^7 - 10^8 $ & $35-37$ \\ \hline
$P_{\delta g}  \, \rightarrow \, {\rm GW}$ @ PTA  & $10^{6} - 10^{8}  $ & $36-40$ \\ \hline
$P_{\delta g}  \, \rightarrow \, {\rm GW}$ @ LISA  & $10^{11} - 10^{14}  $ & $22-28$ \\ \hline
$P_{\delta g}  \, \rightarrow \, {\rm GW}$ @ AdvLIGO  & $10^{16} - 10^{17}  $ & $15-17$ \\ \hline
\end{tabular}
\caption{First column: list of observational windows on inflation. Second column: order of magnitude of the wavenumber of the primordial modes in the corresponding window. Third column: estimated number of efolds before the end of inflation at which those modes exited the horizon. The third, fourth, and fifth row refer to GW produced by the collision of black hole binaries originated by PBH due to enhanced scalar perturbations produced during inflation. For brevity, we denote by AdvLIGO the regime probed by terrestrial GW interferometers. The last three rows denote a stochastic GW signal produced directly during inflation. 
} 
\label{tab:window}
\end{table}
\end{center}

Table \ref{tab:window} summarizes these observational windows, together with the number of e-folds before the end of inflation at which the corresponding modes were generated.\footnote{The figures in the table should be understood as order of magnitude estimates, and have been obtained with some rounding errors. In these figures, we have assumed a constant Hubble rate  $H = 10^{13} \, {\rm GeV}$ during inflation, and that the  Planck pivot scale $k= 0.002 \, {\rm Mpc}^{-1}$ exited the horizon at $N=60$, resulting in $N\simeq 53.8 - \ln (k/{\rm Mpc}^{-1})$. In relating $N$ to the PBH mass we assumed instantaneous thermalization after inflation. From eq. (\ref{MBH-N}), this gives $N \simeq 37.9 + 1/2 \, \ln (M/\Msun)$, where $M_\odot \simeq 2 \times 10^{33} \, g$ is the mass of the sun.} It is fair to say that, while the near scale invariance and the other properites that we observe in the CMB radiation are a general prediction of inflation, the inflationary GW signals discussed in the table do not need to be present, and in fact do not arise in the minimal models characterized by an uncoupled inflaton moving in slow roll. Nonetheless, the unique opportunity offered by this new observational windows highly motivates the study of inflationary models or mechanisms that can provide such signals. 

In this work we focus on one such scenario, namely on the amplification of gauge modes in axion inflation, and on the GW sourced by these modes.\footnote{See Ref.~\cite{Guzzetti:2016mkm} for a review of GW production during inflation from this and from other mechanims.}  Axion inflation (for a review, see \cite{Pajer:2013fsa}) is a very natural class of inflationary models \cite{Freese:1990rb}, as the flatness of the inflaton potential is protected from large quantum corrections by a (softly broken) shift symmetry. The symmetry highly constrains the coupling of the inflaton, and the inflaton decay typically proceeds through the dominant dimension-five operator 
\begin{equation}
\Delta {\cal L} = - \frac{1}{4 f} \, \phi \, F_{\mu \nu} {\tilde F}^{\mu \nu} \;, 
\label{phi-F-Ft}
\end{equation} 
where $\phi$ denotes the inflaton field,  $F$ is the field strength of a gauge field, and ${\tilde F}$ its dual. The quantity $f$ has the dimensions of mass, and it is typically denoted as the axion decay constant. Since Ref.~\cite{Anber:2006xt}, it was realized that, in the presence of this coupling, the motion of the inflaton amplifies only one of the two vector polarizations during inflation. The produced vector modes can then have several phenomenological consequences, including the generation of primordial magnetic fields \cite{Anber:2006xt}, CMB  non-gaussianity \cite{Barnaby:2010vf}, increases scalar power at large scales \cite{Meerburg:2012id}, primordial black holes \cite{Linde:2012bt,Cheng:2016qzb}, chiral gravitational waves~\footnote{While the original formulations of the models \cite{Adshead:2012kp,Maleknejad:2011jw} are ruled out by CMB observations \cite{Dimastrogiovanni:2012ew,Adshead:2013nka,Namba:2013kia}, chiral gravity waves can also be sourced by gauge fields in extensions of Chromo-Natural inflation and Gauge-flation \cite{Maleknejad:2016qjz,Obata:2016xcr,Dimastrogiovanni:2016fuu,Adshead:2016omu}.}  at CMB \cite{Sorbo:2011rz,Barnaby:2012xt,Namba:2015gja} and interferometer \cite{Cook:2011hg,Barnaby:2011qe,Domcke:2016bkh} scales. 

Obtaining an inflationary GW signal at interferometer scales requires either a bump or a blue spectrum, to increase the GW signal from the low value that it has at CMB scales. Such a blue signal is a natural expectation from (\ref{phi-F-Ft}). This term is a total derivative in the case of constant $\phi$ (consistent with the fact that the operator (\ref{phi-F-Ft}) is compatible with the shift symmetry). Therefore, the gauge field amplification is proportional to the speed of the inflaton field, which naturally increases towards the end of inflation. As with signatures on CMB scales \cite{Barnaby:2010vf}, the main issue in models that generate a large GW signal is to make sure that they do not simultaneously overproduce scalar perturbations. For observables generated in the later stages of inflation, such as GW at terretrial interferometers, the main concern is that the scalar perturbations could lead to the overproduction of PBH \cite{Linde:2012bt,Bugaev:2013fya,Erfani:2015rqv,McDonough:2016xvu}. In fact, ref. \cite{Linde:2012bt} showed that, if the existing analytic computations of the scalar perturbations induced from (\ref{phi-F-Ft}) at small scales are accurate, the PBH limits prevent the gauge field amplification to be strong enough to generate a visible GW signal at LISA and AdvLIGO in models of chaotic field inflation.  The main purpose of the present work is to understand whether and under which conditions this conclusion can be evaded. The first observation,  is that there is an $\calO$ uncertainty on the amount of scalar perturbations generated by this mechanism in the regime which is necessary to produce a large enough signal. As already pointed out in Ref.~\cite{Linde:2012bt}, an $\calO$ decrease of the scalar signal would be enough for relaxing the PBH bound to allow a visible GW signal. Here we stress two additional possibilities. 

Firstly, we note that, due to the typical blue spectrum of the sourced perturbations, the PBH limit is enforced by modes at much smaller scales than those probed by interferometers, particularly for the LISA and PTA cases: the limit is enforced by modes generated at $N \simeq 10$, while LISA (resp, PTA experiments) is mostly sensitive to modes produced at $N \sim 25$ (resp. $N \sim 40$) before the end of inflation. Without committing to any specific inflationary potential (which is a necessity, if one wants to relate signatures at different scales), we ask the question whether the PBH limit at any given scale precludes an observable GW signal  at that scale. The question in nontrivial, and it must be answered by an explicit computation, as the answer ultimately depends on the sensitivity of the GW measurement. Our computations provide a positive answer for the PTA-SKA projected sensitivity  \cite{Moore:2014lga} and for the  LISA sensitivity curves reported in \cite{Bartolo:2016ami}, and a negative one for the current and projected AdvLIGO sensitivity curves reported in \cite{TheLIGOScientific:2016wyq}. We reached this conclusion for two different implementation of the mechanism. The first one is a modification of the inflaton potential short after the LISA (or AdvLIGO) modes are produced; specifically, we arranged for a fast decrease of the inflaton slope (in the concrete example we studied, we showed that a decrease of the slope of a factor of $3$ is sufficient to slow down the inflaton, and to sufficiently weaken the PBH constraint). The second model we considered is a two field model, recently introduced in Ref.~\cite{Namba:2015gja} to produce a localized bump in the GW spectrum at CMB scales. In this model, the gauge field amplification is due to a pseudoscalar $\sigma$ that is not the inflaton, and that rolls only for a few ($\sim 5$) e-folds of inflation.\footnote{This requires tuning the mass of the field to $m = {\cal O }(H)$, as well as tuning the initial conditions for $\sigma$. We do not explore in detail these aspects of the model, since we only use this toy model as a proof of principle to test the generality of the conclusions reached in the $\phi \, F {\tilde F}$ case.} Also this implementation leads to a positive (negative) conclusion for the projected PTA-SKA and LISA (AdvLIGO) sensitivity curves.

Secondly, the conclusion of  Ref.~\cite{Linde:2012bt} can be also evaded if the coupling (\ref{phi-F-Ft}) involves a set of ${\cal N} > 1$ gauge fields (as would be the case, for instance, with a non-abelian gauge group). The presence of more gauge fields has several consequences that we describe in our analysis. One of them is that the  ratio between the tensor and scalar power spectrum increases as ${\cal N}^2$ in the regime of significant gauge field amplification. Based on this, one could expect that even a moderate value of ${\cal N}$ can result in a visible GW signal, without exceeding the PBH bound. Ref. \cite{Domcke:2016bkh} showed that this is indeed possible at  ${\cal N} = 10$ in the case of Starobinsky inflation \cite{Starobinsky:1980te}. We show that this can be achieved with even fewer fields, ${\cal N } = 6$, in the case of chaotic inflation. This is a very moderate value for a unified gauge group.

This discussion refers to the GW signal directly sourced by the gauge fields during inflation, and that gives rise to the observational windows listed in the last two rows of Table \ref{tab:window}. As we mentioned, the mechanism of gauge field amplification also gives rise to a different, and more complicated, mechanism of GW production.  Specifically, it may be possible that the scalar perturbations give rise to a significant amount of PBH compatible with observations, that act as seeds for structure and may be responsible for most of the observed non-stellar black holes. These PBH could also be identified as the dominant component of the dark matter of the universe~\cite{Clesse:2015wea,Bird:2016dcv,Clesse:2016vqa,Sasaki:2016jop,Kashlinsky:2016sdv,Kawasaki:2016pql}. The evolution from the seeds to the present black holes involves merging of the different PBH and gas accretion~\cite{Clesse:2016vqa}, and is beyond the scope of the present work. Nevertheless, we briefly review this possibility here, since the gauge field amplification provides a mechanism for the generation of the seeds which is alternative to another mechanism that have been proposed in the context of hybrid inflation~\cite{GarciaBellido:1996qt,Clesse:2015wea}. We discuss this possibility in the context of the two-field $\phi-\sigma$ model, as it produces a localized bump of scalar field modes, and therefore a narrow spectrum of black hole masses, for which estimates can be more easily made. As we have already mentioned, this new mechanism of GW production provides a different window on inflation (third and fourth rows of Table \ref{tab:window}). 

The plan of the paper is as follow. In Section \ref{sec:PBG-GW} we review the limits on PBH and on the scalar perturbations generated from this mechanism. In Section \ref{sec:phi} we discuss the GW production for the direct coupling $\phi\, F {\tilde F}$ between the inflaton and the gauge field. In Section \ref{sec:sigma} we discuss the case of localized GW production from a field $\sigma$ different from the inflation. Section \ref{sec:conclusions} provides a summary of our results and some concluding remarks.

\section{Limits on scalar perturbations, and prospects for the detection of a stochastic GW background}
\label{sec:PBG-GW}

\begin{figure}[tbp]
\centering 
\includegraphics[width=0.45\textwidth,angle=0]{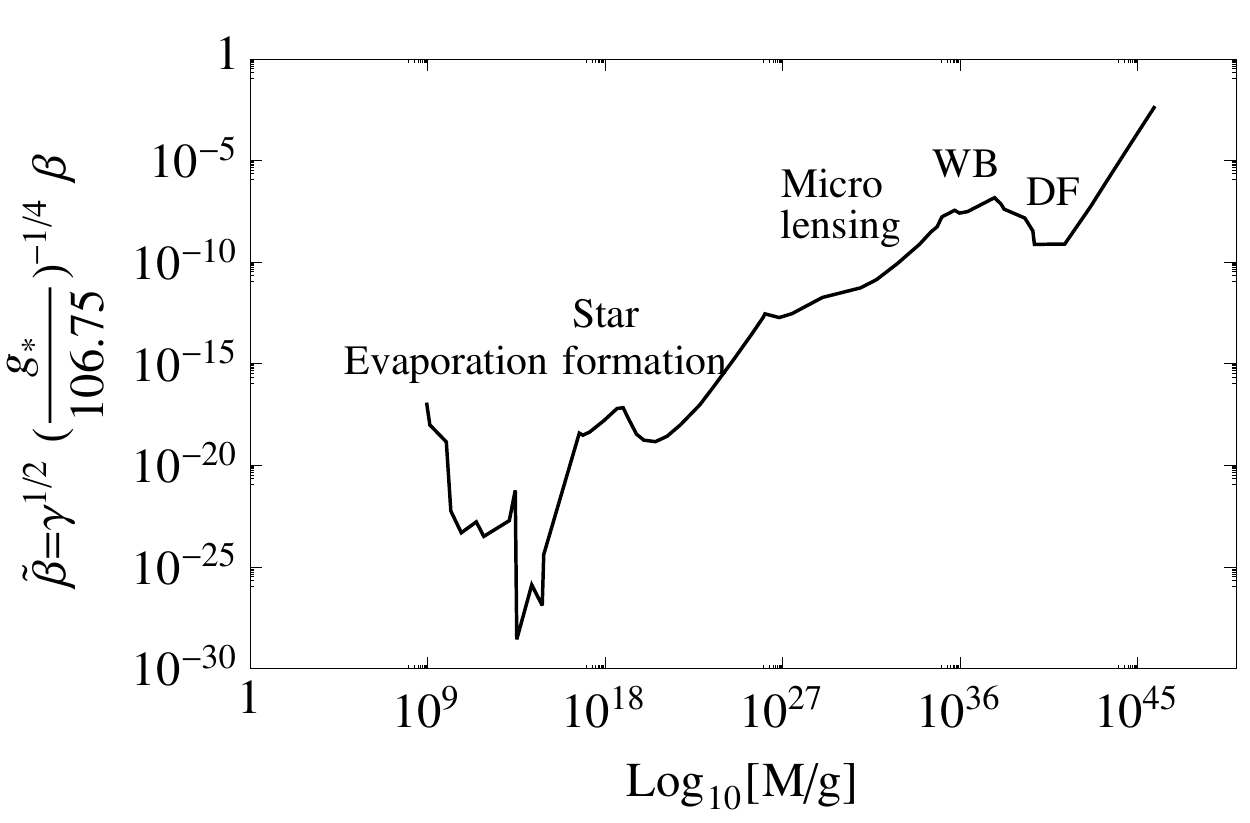}
\includegraphics[width=0.45\textwidth,angle=0]{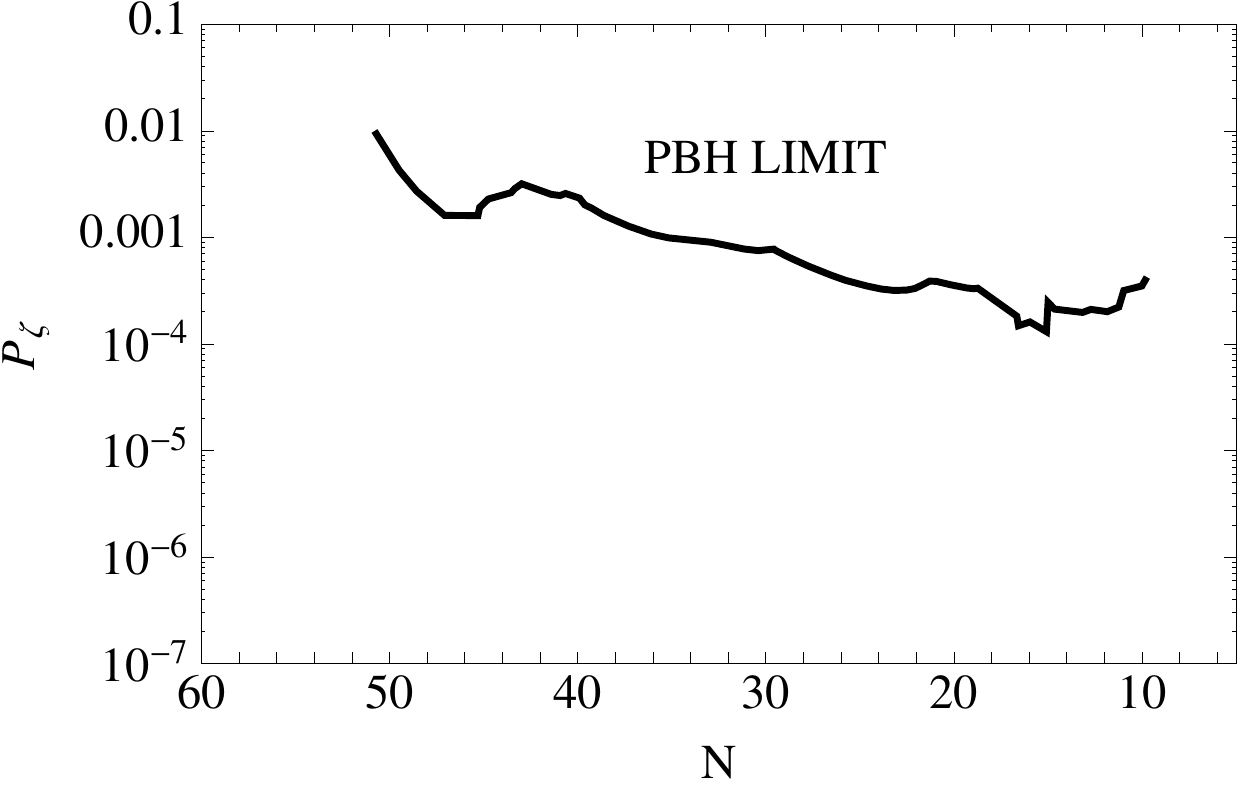}
\hfill
\caption{ 
Left panel: Limits on the rescaled black hole fraction $\betap$ as a function of the black hole mass, as discussed in the text. 
Right panel: Same limits, written as a bound on the primordial scalar density perturbations as a function of number of e-folds before the end of inflation, assuming a constant Hubble rate $H = 10^{13} \, {\rm GeV}$ during inflation, and a $\chi^2$ statistics of the scalar perturbations (see Appendix \ref{app:zetaPBH}). 
}
\label{fig:limits}
\end{figure}

In the left panel of Fig.~\ref{fig:limits}, we present the limits on PBH considered in this work. We show the limits in terms of the rescaled variable $\betap$ introduced in \cite{Carr:2009jm} (where it was denoted as $\beta'$) as a function of the black hole mass. As we explain in Appendix \ref{app:zetaPBH}, the quantity $\betap$ is related through equation (\ref{betatilde-beta})  to the fraction of regions (of a given size, corresponding to a given black hole mass)  that collapse to form a black hole $\beta$. We now list the origin of the limits included in this figure. 

Going from smaller to greater black hole mass $M$,\footnote{We do not  include limits at $M < 10^9 g$ shown in Fig.~9 of  \cite{Carr:2009jm} as they are model dependent, and assume that the black hole evaporation leaves behind stable relics.} the first limits shown in Fig.~\ref{fig:limits} are a consequence of the black hole evaporation which, depending on $M$, can photodissociate elements formed during Big-Bang Nucleosynthesis, modify the CMB, distort the galactic and extra-galactic $\gamma-$ray background.   This set of constraints is studied in detals in Ref.~\cite{Carr:2009jm} (see also \cite{Khlopov:2008qy}). The constraints we show in the mass range $5 \times 10^{16} \la M({\rm g}) \la 10^{26}$ are obtained from the effects of the capture of black holes by stars (the black holes would eventually destroy the neutron star or white dwarf remnants),\footnote{These updated constrains are stronger than those arising from the lack of observation of femtolensing of $\gamma-$ray bursts \cite{Barnacka:2012bm}, and the lack of microlensing events at Kepler \cite{Griest:2013aaa}, which we therefore do not show here.} according to the updated computations of Ref.~\cite{Capela:2014ita}. The limits in the  $10^{26} \la M({\rm g}) \la 10^{35}$ range are due to the lack of observations of short duration microlensing events by the MACHO and EROS Collaborations~\cite{Tisserand:2006zx,Alcock:1998fx}. These experiments lasted for approximately six years each, and thus could not constrain higher masses of Massive Compact Halo Objects, which correspond to long duration events of order a decade. The limit indicated as ``WB'' is obtained from the non observation of wide binary disruption \cite{Quinn:2009zg}. Finally, the ``DF'' limit refers to dragging of halo objects into the Galactic nucleus by dynamical friction \cite{Carr:1997cn}.  For the last three constraints, we took the limits as reported in Figure 3 of  \cite{Carr:2016drx}. We did not include the CMB bounds from $y-$ and $\mu-$distortions~\footnote{$y-$ and $\mu-$distortions also put a bound on the scalar power spectrum (independently on whether PBH form) for modes that re-enter the horizon at $ z \la 10^6$, due to the fact that the energy associated with these modes does not perfectly thermalizes with the background energy density. $\mu-$distortions are mostly affected by primordial perturbations of wavenumbers  $50\, {\rm Mpc}^{-1}\la k \la10^4\, {\rm Mpc}^{-1}$ \cite{Chluba:2011hw,Chluba:2012we}, which roughly corresponds to modes that left the horizon at $45 \la N \la 50$ e-folds before the end of inflation (as in the rest of the paper, this assumes that  the Planck pivot scale $k=0.002 \, {\rm Mpc}^{-1}$ corresponds to $N=60$.)  $y-$distortions are instead mostly sensitive to modes  $1\, {\rm Mpc}^{-1}\la k \la 50\, {\rm Mpc}^{-1}$  \cite{Chluba:2011hw,Chluba:2012we},  which roughly corresponds to $50 \la N \la 54$. All the sourced signals that we consider in this work take place at scales much smaller than these, and therefore do not induce these distortions.}  due to X-rays emitted by gas accretion onto PBHs before recombination, because there was an error in Ref.~\cite{Ricotti:2007au} and the bounds have shifted by several orders of magnitude,\footnote{Jens Chluba and Yacine Ali-Ha\"imoud, private communication.} beyond WB bounds, where DF bounds are stronger. We also ignored the Eridanus-II bounds of Ref.~\cite{Brandt:2016aco} because the star cluster at the center of the dwarf galaxy could be stabilized by an intermediate mass BH, which would prevent the puffing up of the system. This effect shifts the bounds again below the DF bounds.

In Appendix ~\ref{app:zetaPBH} we show how these limits translate into an upper bound (which we show in the right panel of Figure \ref{fig:limits}) on the amount of primordial scalar perturbations $P_\zeta$ as a function of the number of e-folds $N$ before the end of inflation at which they leave the horizon.  Our computations focus on the PBH formed by scalar perturbations caused by gauge field amplification in axion inflation  \cite{Linde:2012bt}. The scalar modes obey a $\chi^2$ statistics   \cite{Linde:2012bt}, which significantly tightens the limit on $P_\zeta$ with respect to Gaussian statistics, as we show in Appendix  \ref{app:zetaPBH}. We also show there that,  in the relevant range, $\beta$ changes dramatically with $P_\zeta$, which explains why the limit in the left panel of Figure \ref{fig:limits} changes by $\sim 25$ orders of magnitude, while that in the right panel changes by only $\sim 2$. 

The field amplification that enhances the scalar perturbations also produces a primordial stochastic background of gravitational waves. 
The goal of this work is to study under which condition this background can be observed, without violating the bounds from PBHs. 

It is customary to express the  amplitude of the GW background in terms of their present fractional energy density  per logarithmic wavenumber interval $\Omega_{\rm GW}$, which is related to the GW power spectrum by (see \cite{Barnaby:2011qe} for  details) 
\begin{equation}
\Omega_{\rm GW} \equiv \frac{1}{3 H_0^2 M_P^2} \, \frac{\partial \rho_{\rm GW,0}}{\partial \log \, k} = \frac{\Omega_{\rm R,0}}{24} \left( P_L + P_R \right) \;, 
\label{OmGW}
\end{equation} 
where $\Omega_{{\rm R},0} \, h^2 \simeq 4.2 \times 10^{-5}$ refers to radiation today (including neutrinos, as if they were still relativistic). 
The quantity $\Omega_{\rm GW}$ is typically plotted as a function of frequency $f = k / 2 \pi$, which is related to the number of e-folds by
\begin{equation}
N  = N_{\rm CMB}  - 44.92 + {\rm ln } \left( \frac{k_{\rm CMB}}{0.002 \, {\rm Mpc}^{-1}} \right) - {\rm ln } \left( \frac{f}{100 \, {\rm Hz}} \right) + 
{\rm ln } \left( \frac{H_N}{H_{\rm CMB} } \right) 
\label{N-f}
\end{equation} 
where $N_{\rm CMB}$ is the number of e-folds at which the mode $k_{\rm CMB}$ left the horizon. In (\ref{N-f}) we have normalized this scale to the Planck pivot scale, which in this work we assume to correspond to $N_{\rm CMB} = 60$. We note that the last term in (\ref{N-f}) accounts for the variation of the Hubble rate during inflation ($H_N$ denotes the value at $N$ e-folds before the end of inflation, and $H_{\rm CMB} $ at $N=N_{\rm CMB}$). This factor was  neglected in Ref.~\cite{Barnaby:2011qe}. 

In the following sections we compare the GW signal produced in models of axion inflation with the sensitivity curves of  AdvLIGO (from top to bottom, O1-O3-O5 lines of Figure 1 of \cite{TheLIGOScientific:2016wyq}), LISA (from bottom to top,  A5M5-A5M2-A2M5-A2M2-A1M5-A1M2  lines of Figure 1 of \cite{Bartolo:2016ami}; the various lines refer to different choices for the length of the LISA arms and for the LISA duration. Specifically, the labels A1, A2, and A5 correspond, respectively, to 1, 2, and 5 million km; the labels M2 and M5, correspond, respectively, to 2 and 5 years), and of Pulsar Timing Array experiments (the curve labeled by PTA corresponds to the combination of the current limits of 
\cite{Arzoumanian:2015liz,Lentati:2015qwp,Shannon:2015ect}; the curve labeled by SKA corresponds to the forecast sensitivity of the Square Kilometer Array PTA experiment, obtained using the GWPlotter tool \cite{Moore:2014lga}).

\section{Production from a rolling inflaton $\phi$  } 
\label{sec:phi}

In this section we  study the phenomenology of the inflationary model 
\begin{equation} 
{\cal L} = - \frac{1}{2} \left( \partial \phi \right)^2 - V \left( \phi \right) - \frac{1}{4} F^2 - \frac{\phi}{4f} \, F \, {\tilde F} \;, 
\label{lagrangian-phi} 
\end{equation} 
for a pseudo-scalar inflaton $\phi$ coupled to a U(1) vector field ($F$ is the field strength associated to the vector, and ${\tilde F}$ its dual). We divide the discussion in two subsections. In the first one, we review results from the literature to compute the scalar and tensor (GW) primordial perturbations generated in this model. In the second subsection we then show that, depending on the inflaton potential and on the coupling strength to vector, this model can result in observable gravity waves at interferometer scales, without overproducing scalar perturbations  and primordial black holes.

\subsection{Production of scalar and tensor modes }
\label{subsec:phi-part1}

We consider inflation in the model described by the lagrangian (\ref{lagrangian-phi}). Due to the motion of the inflaton, the coupling amplifies one circular polarization of the vector field, leading to \cite{Anber:2006xt} 
\begin{equation} 
A_+ \simeq \frac{1}{\sqrt{2 k}} \left( \frac{- k \tau}{2 \xi} \right)^{1/4} \! \exp\left(\pi \xi - 2 \sqrt{-2\xi k \tau}\right)\,,
\hspace{2cm} \xi \equiv \frac{\dot\varphi}{2 f H} \;, 
\label{Apsol} 
\end{equation} 
where $k$ is the comoving momentum of the gauge field mode, $\tau$ is conformal time, and $H$ the Hubble rate (we are assuming $\xi > 0$; in the opposite case, the $A_-$ polarization is amplified). Moreover, $\varphi$ denotes the zero mode of the inflaton field, $\phi = \varphi(t) + \delta \phi \left( t ,\, \vec{x} \right)$.  The amplification of any gauge field mode  takes place only when the size of the mode is comparable to the horizon, so that this mechanism has no UV nor IR singularities: when deep inside the horizon, the gauge field mode has a standard dispersion relation, and it is not amplified; at very large scales, the growth weakens and it becomes subdominanat to the expansion of the universe, so that the gauge field is diluted away by the expansion. The approximate expression (\ref{Apsol}) is valid in the interval $1/8 \xi \ll - k \tau \ll 2 \xi$. These are the times during which the mode grows, and remains at a sizable level, before being diluted away. Therefore this approximation allows to obtain a good estimate of the phenomenological signatures caused by the mode.\footnote{We omit several details of the computations in this work. We refer the interested reader to Section 2.1 of \cite{Peloso:2016gqs} for a detailed summary of the gauge field amplification; to Section II.B of \cite{Barnaby:2011qe} for the study of the backreaction of the vector field on the background dynamics; to Section IV A of \cite{Barnaby:2011vw} and Section II of  \cite{Linde:2012bt} for the computation of the scalar perturbations sourced by the vector modes during inflation; finally, to Section V of  \cite{Barnaby:2011vw} and to Section V of  \cite{Barnaby:2011qe} for the  computation of the GW sourced by the vector modes during inflation.} 

The gauge modes source scalar perturbations (through the inverse decay $A_+ + A_+ \rightarrow \delta \phi$ process) and gravitational waves (through the gravitational interaction $A_+ + A_+ \rightarrow h_+$, and  $A_+ + A_+ \rightarrow h_-$,  where $h_\pm$ denotes the two GW polarizations). These signals add up incoherently to the usual vacuum ones, so that the scalar and tensor correlators are the decoupled sum of a vacuum (which we denote by the suffix v) and sourced (suffix s) contribution.  The fact that only $A_+$ modes are amplified (which is due to the breaking of parity associated to the pseudo-scalar inflaton) results in a much greater production of $h_+$ with respect to $h_-$, namely the sourced GW background breaks parity nearly maximally \cite{Sorbo:2011rz}. In the following, we disregard the highly subdominant sourced $h_-$ mode. One finds~\footnote{The two functions in this parametrization have the limiting behavior  ${\hat f}_{2,\zeta} \left( \xi \right) \simeq 
7.5  \times 10^{-5}/\xi^6$ and ${\hat f}_{2,+} \left( \xi \right) \simeq 4.3  \times 10^{-7}/\xi^6$  in the $\xi \gg 1$ regime. The full dependence is given in  \cite{Barnaby:2011vw}.  We note the presence of a typo in eq. (3.40) of  \cite{Barnaby:2011vw}, namely the r.h.s. should contain an additional $\frac{1}{4}$ factor. This typo did not propagate to any of  the other equations or figures of that work.}
\begin{eqnarray}
P_{\zeta} \left( k \right) &\simeq&   P_{\zeta,v} \left( k \right) +  P_{\zeta,v}^2 \left( k \right) \, {\hat f}_{2,\zeta} \left( \xi \right) \, {\rm e}^{4 \pi \xi} \;\;, \nonumber\\ 
P_{\rm GW} \left( k \right) &=& P_{GW,+} \left( k \right) +  P_{GW,-} \left( k \right) \simeq \frac{2 H^2}{\pi^2 M_p^2} \left[ 1 + \frac{H^2}{M_p^2} \, {\hat f}_{2,+} \left( \xi \right) \, {\rm e}^{4 \pi \xi} \right] \;, 
\label{Pz-PGW}
\end{eqnarray} 
where we recall the standard vacuum result $ P_{\zeta,v}  \simeq \
H^4/(4 \pi^2 \dot{\varphi}^2)$. 

Typically, the parameter $\xi$ adiabatically grows during inflation (we note that $\xi \simeq \sqrt{\frac{\epsilon}{2}} \, M_p/f$, where $\epsilon$ is the usual slow roll parameter). Therefore,  the quantity $\xi$ in eqs.  (\ref{Apsol}) and  (\ref{Pz-PGW}) should be understood as the value acquired by $\xi$ when the mode in consideration crosses the horizon  (the same is true for the quantity $H$ appearing in (\ref{Pz-PGW})).  At CMB scales, $\xi$ is constrained by non-gaussianity \cite{Barnaby:2010vf,Ade:2015lrj} and by the growth of the scalar spectrum with $k$ \cite{Meerburg:2012id,Ade:2015lrj} (one typically obtains $\xi_{\rm CMB} \la 2.2 - 2.5$, depending on the specific potential, and priors \cite{Ade:2015lrj}). For such values of $\xi$, the amplified gauge quanta modify the background dynamics in a negligible manner. 

As the inflaton speeds up during inflation, $\xi$ increases.  At larger values of $\xi$, the amplified gauge modes can significantly backreact on the background evolution. The dominant effect is an additional friction in the equation for the inflaton field \cite{Anber:2009ua} (the physical reason for this is that the gauge field amplification occurs due to the motion of $\varphi$, and therefore at the expense of the inflaton kinetic energy). This can give rise to a transition \cite{Barnaby:2011qe} between the usual slow roll evolution at early times, and a new attractor background solution at late times, where the gauge field amplification dominates  over the Hubble friction. The increase of $\xi$ also gives rise to a significant increase of the sourced inflaton perturbations and gravity waves. 

The increase of the inflaton perturbations complicates the system of scalar perturbations. Eq. (\ref{Apsol}) for the vector modes has been obtained for a homogeneous inflaton (we note that $A_+ \left[ \xi \left[ \varphi \right] \right]$ in that expression). However, we expect that at sufficiently large $\xi$ the inflaton perturbations will be large enough as to significantly impact the gauge mode solution, so to go from the solution  (\ref{Apsol}) to a more general $A_+ \left[ \varphi + \delta \phi \right]$. This quantity acts as a source in the equation for the inflaton perturbations. Expanding this source in $\delta \phi$ introduces additional terms in the equation for the inflaton perturbations, which must be relevant at sufficiently high $\xi$.  A complete equation for the scalar perturbations in this regime has not yet been obtained. Here we employ the approximate  analytic equation first derived in \cite{Anber:2009ua}, and later used in \cite{Barnaby:2011qe,Linde:2012bt}~\footnote{Following standard notation, dot denotes derivative with respect to physical time, while prime on a function denotes derivative with respect to its argument.} 
\begin{equation}
\delta \ddot{\phi} + 3 \beta H \delta \dot\phi + k^2 \delta \phi + V'' \delta \phi = \frac{\vec{E} \cdot \vec{B} - \langle \vec{E} \cdot \vec{B} \rangle}{f} \;, 
\label{eq-deltaphi}
\end{equation}
where 
\begin{equation}
\beta \equiv 1 - \frac{2 \pi \xi}{f} \, \frac{\langle \vec{E} \cdot \vec{B} \rangle}{3 H \dot\varphi}\,. 
\label{beta-lorenzo}
\end{equation}
In our expressions, $\vec{E}$ and $\vec{B}$ denote the ``electric'' and ``magnetic'' field associated with the solution (\ref{Apsol}).\footnote{We use standard electromagnetic notation for convenience, but we are not necessarily implying that the vector field in (\ref{lagrangian-phi}) is the Standard Model photon.} Taking $\beta =1$ in (\ref{eq-deltaphi}) amounts in neglecting any dependence of the vector fields on $\delta\phi$. The sourced scalar solution in (\ref{Pz-PGW}) has been obtained with this assumption, and, for the reasons we just discussed we expect it to be accurate at sufficiently small $\xi$. Taking instead $\beta$ as in (\ref{beta-lorenzo}) amounts into including one of the effects arising from the dependence of the gauge field from $\delta\phi$. As the solution (\ref{Apsol}) depends on the time derivative of the inflaton zero mode, it is reasonable to expect that the first effect of $\delta \phi$ in the source will be through the change $\delta \xi = \frac{\delta \xi}{\delta \left( \delta\dot{\phi} \right)} \delta\dot{\phi}$, which precisely leads to (\ref{beta-lorenzo}). This was the reasoning adopted in \cite{Anber:2009ua,Barnaby:2011qe,Linde:2012bt} and we also follow it in the present work. However, we alert the reader that eq. (\ref{eq-deltaphi}) is not a complete one, and that additional terms, not included in this equation, may also become important when the departure of $\beta$ from $1$ becomes relevant. 

As shown in \cite{Barnaby:2011qe,Linde:2012bt}, eq. (\ref{eq-deltaphi}) leads to the following estimate for the scalar power spectrum 
\begin{equation}
P_{\zeta,s} \left( k \right) \simeq \left(  \frac{\langle \vec{E} \cdot \vec{B} \rangle}{3 \beta H \dot{\varphi} \, f} \right)^2 \, {\cal F}^2 \,. 
\label{Pz2} 
\end{equation} 
As seen for instance from Figure 4 of \cite{Linde:2012bt} the sourced scalar solutions in (\ref{Pz-PGW}) and (\ref{Pz2}) are in remarkable agreement with each other at sufficiently small $\xi$. This is in particular true for the values of $\xi$ at which the CMB bounds have been derived. However, as $\xi$ increases, the solution (\ref{Pz2}) becomes significantly smaller than that in  (\ref{Pz-PGW}), due to the additional friction included in (\ref{eq-deltaphi}) for $\beta \neq 1$. In the results that we present below, the sourced scalar spectrum (\ref{Pz2}) is used. 

The factor ${\cal F}$ in the expression (\ref{Pz2}) was not present in the computations of refs.  \cite{Anber:2009ua,Barnaby:2011qe,Linde:2012bt}. 
This factor accounts for the impact of the energy density in the gauge field to the denominator of $\zeta \equiv - \frac{H \delta \rho}{\dot{\rho}}$. 
We evaluate it in Appendix \ref{app:calF}, following the observations recently made in \cite{Notari:2016npn}. Although we agree with 
 \cite{Notari:2016npn} that this factor should be present, we find that it has only a marginal relevance. From both numerical and analytical computations, we find that this factor evolves from $1$ in the regime of negligible backreaction of the gauge field on the background dynamics (appropriate at CMB scales)  to $\frac{7}{8}$ in the regime of strong backreaction (appropriate of late times). The limiting value $\frac{7}{8}$ is independent of the inflaton potential,  of the value of $f$, and of the number of amplified gauge fields. In the evolutions studied in this paper, this limiting value is reached only at the very end of inflation.  To give a reference value, ${\cal F} \simeq 0.98$ at $N=10$ in the evolution shown in Figure \ref{fig:lin-1p66}. 

The gauge fields also source metric perturbations, and the solution (\ref{Apsol}) has been used in obtaining the second line of (\ref{Pz-PGW}). Metric perturbations would modify  (\ref{Apsol}) in a completely negligible manner (just due to gravity). Therefore we can ignore any $\frac{\delta A_+}{\delta h} \, \delta h$ correction in the source of the equation for the gravitational waves, and use the second line of (\ref{Pz-PGW}) in the results presented below.

\subsection{Phenomenology at sub-CMB scales }
\label{subsec:phi-part2}

We are now ready to discuss the phenomenology of this model at scales much smaller than the CMB ones. In particular, we want to review the result of Ref.~\cite{Linde:2012bt}, which appears to be a major obstacle against the possibility of observing GW  at interferometer scales from the model (\ref{lagrangian-phi}).  Ref.~\cite{Linde:2012bt} computed the primordial scalar and tensor perturbations produced in this model for a quadratic inflaton potential. They found that, due to the non-Gaussian nature of the scalar perturbations, the scalar field amplification can easily overcome bounds imposed by PBH. They can be avoided only provided that $\xi_{\rm CMB} \la 1.5$ at $N=60$ e-folds before the end of inflation. Once this limit is respected, the GW production is below the PTA-SKA, LISA, and AdvLIGO sensitivities. As by now the quadratic inflaton potential is ruled out by the CMB \cite{Ade:2015lrj}, we perform an analogous computation for a linear inflaton potential, which is motivated by monodromy \cite{Silverstein:2008sg,McAllister:2008hb}.

\begin{figure}
\centerline{
\includegraphics[width=0.35\textwidth,angle=0]{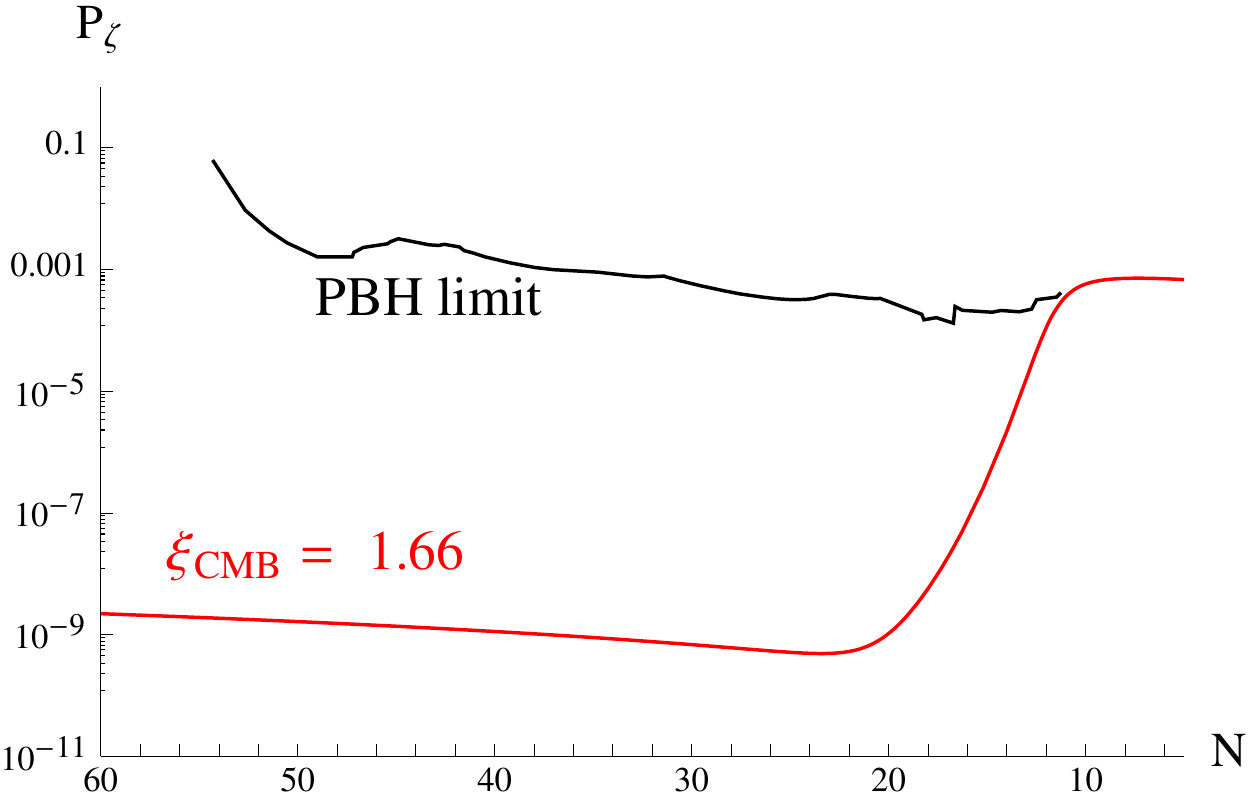}
\includegraphics[width=0.35\textwidth,angle=0]{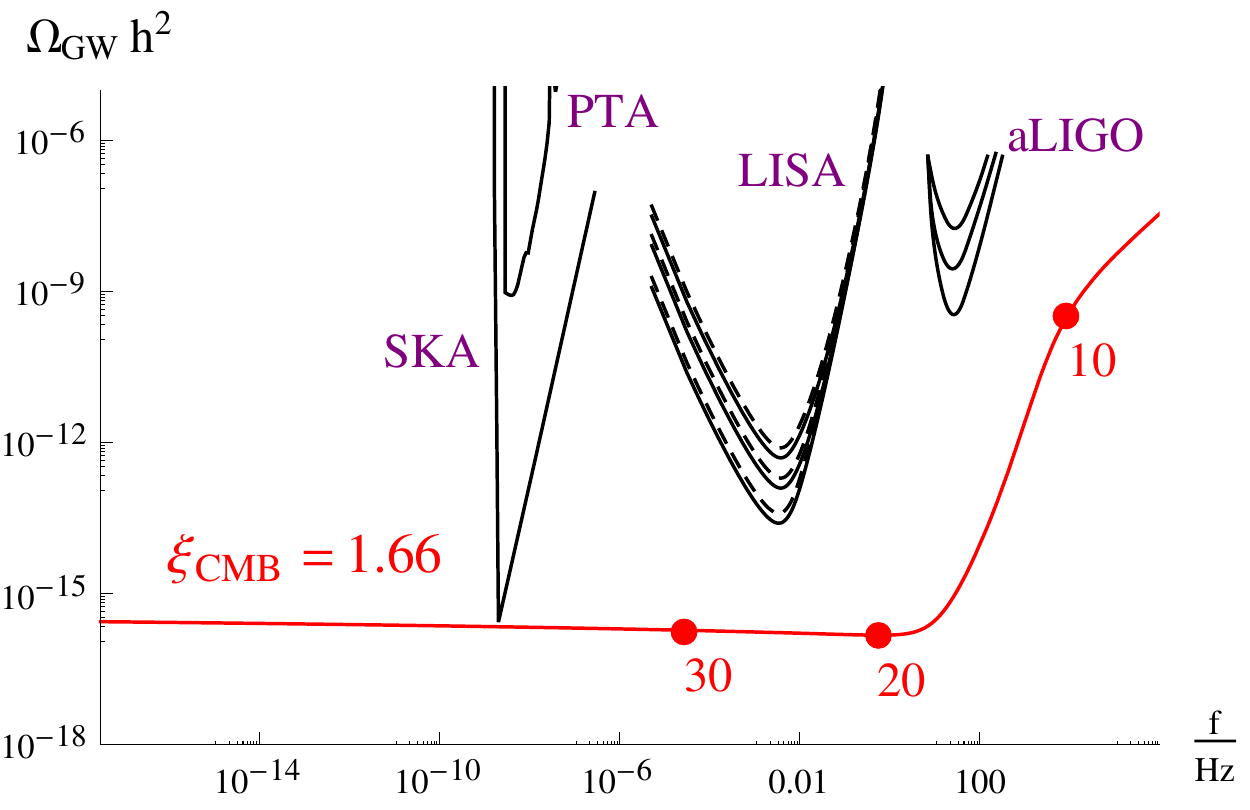}
}
\caption{LEFT PANEL: Scalar power spectrum in axion inflation for a linear inflaton potential, in the approximation (\ref{Pz2}). The coupling to gauge fields is chosen as large as allowed by the PHB bounds shown in the figure; this gives $\xi_{\rm CMB} \la 1.66$ at $60$ e-folds before the end of inflation. RIGHT PANEL: Corresponding GW signal for the same model and parameters as in the left panel. The points on the theoretical line labelled by ``$30$'',  ``$20$'', and  ``$10$'', correspond to modes that, respectively, left the horizon $30, 20,$ and $10$ e-folds before the end of inflation. 
}
\label{fig:lin-1p66}
\end{figure}

We numerically integrate the background equation of motion for the inflaton and the Friedmann equation, keeping into account the backreaction of the produced gauge quanta in both equations \cite{Barnaby:2011qe}. We iteratively vary the axion scale $f$ and the mass parameter in the linear potential so to obtain the correct power spectrum normalization $P_\zeta = 2.2 \times 10^{-9}$ \cite{Ade:2015lrj} at $N=60$, and so to obtain the largest inflaton-vector coupling (smallest $f$) allowed by the PBH limit. This results in $\xi_{\rm CMB} \simeq 1.66$ at $N=60$,  close to the value found in   \cite{Linde:2012bt} for a quadratic inflaton potential.\footnote{We note that the upper bound on $\xi$ is obtained from the latest times (namely, from the smallest values of $N$) for which the PBH limit is present. This corresponds to the smallest PBH masses $M$ for which the limit is present, see eq. (\ref{MBH-N}). (i) We only consider limits for $M > 10^9 g$, for the physical reasons that we stated in footnote 1. Ref.  \cite{Linde:2012bt} made the same physical assumption, but it considered masses starting from $M > 10^8 g$ (we believe that the $10^9 g$ figure is more in line with Fig.~9 of  \cite{Carr:2009jm}). (ii) We included an efficiency factor $\gamma \simeq 0.2$ in our relation (\ref{MBH-N}). (iii) Our relation (\ref{MBH-N}) accounts for the variation of the Hubble rate during inflation. All these three factors contribute to increase the value of $N_{\rm min}$ at which the PBH bound is present, giving a weaker constraint on $P_\zeta$ in our case, with respect to  \cite{Linde:2012bt}. We repeated their computation for quadratic inflaton potential, but with our limit on $P_\zeta$, obtianing $\xi_{\rm CMB} \la 1.82$, rather than their result $\xi_{\rm CMB} \la 1.5$. This slightly weaker limit does not change the physical conclusion of   \cite{Linde:2012bt}  that the PBH bound prevents GW from being observable in a quadratic inflaton potential.}  We show the scalar spectrum, and the PBH bound, in the left panel of Fig.~\ref{fig:lin-1p66}.  The right panel of the figure shows the GW produced for the same linear potential and parameters choice. We see that, also in this case, the limit on the gauge field amplification imposed by the PBH bound forces the GW signal to be too small to be observed at PTA-SKA, LISA, and AdvLIGO  scales. 

We stress that this conclusion strongly relies on the scalar modes being accurately described by Eq.~(\ref{eq-deltaphi}) in the $\xi \gg 1$ regime. As we discussed after Eq.~(\ref{eq-deltaphi}), this equation should receive additional corrections in this regime, that can change the result (\ref{Pz2})  by an $\calO$ factor. As already remarked in Ref.~\cite{Linde:2012bt}, an  $\calO$ decrease of the scalar power spectrum would be enough to make the  PBHs limit unimportant. Ref. \cite{Ferreira:2015omg} proposed some conditions for the validity of perturbative computations of the scalar perturbations in this model. The conditions were reanalyzed in Ref.~\cite{Peloso:2016gqs}, which showed that these criteria are satisfied for $\xi \la 4.8$. This is parametrically close to the values necessary to generate PBH, indicating that, while Eq.~(\ref{eq-deltaphi}) very likely provides a correct estimate for the amplitude of the scalar modes,  $\calO$ corrections are a possibility. 

If we assume that the result (\ref{Pz2}) is reliable, we can still imagine two simple possibilities that make the PBH limit less important from the specific case studied in  \cite{Linde:2012bt}. We discuss these two possibilities in the two separate parts in the reminder of this subsection.

\subsubsection{Dependence on the inflaton potential} 
\label{subsubsec:phi-part2A}

We observe from Fig.~\ref{fig:lin-1p66}  that the PBH limit $\xi_{\rm CMB} \la 1.66$ is enforced by modes that left the horizon at  $N \simeq 10$ e-folds before the end of inflation. On the other hand, the PTA, LISA, and AdvLIGO bands include modes that left the horizon earlier. This is particularly true for PTA and LISA, which are mostly sensitive to modes produced, respectively, at $N \simeq 40$ and at $N \simeq 25$. Therefore, the interplay between the PBH limit and the possible GW signal strongly depends on the evolution of $\xi$, and, ultimately, on the inflaton potential, in the latest stages of inflation. We have very little knowledge of the inflaton potential after the CMB and LSS modes are produced, and we can easily imagine potentials for which $\xi$ has a peak at some intermediate value of $N$, and then decreases, without ever violating the PBH bounds. 

This is for example immediately achieved if the linear potential  changes its slope at some given point during inflation, as in the Starobinsky model \cite{Starobinsky:1992ts}. In that work, the potential changes its slope abruptly at some given value $\phi = \phi_*$. Here, we consider a ``regularized'' version of that model, that is the smooth transition
\begin{equation}
V = \left\{ \begin{array}{l} 
- M^3 \, \phi  \;\;\;\;\;\;\;\;\;\;\;\;\;\;\;\;\;\;\;\;\;\;\;\;\;\;\;\;\;\;\;\;\;\;\;\;\;\;\;\;\;,\;\;\; \phi < \phi_1 \\ 
M^3 \, \frac{ \left( 1 - r \right) \left( \phi + \phi_1 \right)^2-2 \phi \left( \phi_1 - 2 r \phi_1 + \phi_2 \right)}{2 \left( \phi_2 - \phi_1 \right)} \;\;\;\;,\;\;\; \phi_1 < \phi < \phi_2 \\ 
- M^3 \frac{\left(1-r\right)\left(\phi_1+\phi_2\right)}{2} - r \, M^3 \, \phi  \;\;\;\;\;\;\;\;\;\;\;,\;\;\; \phi > \phi_2 
\end{array} \right. \,.  
\label{modified-Vphi}
\end{equation}  
Namely, we choose the potential to be linear both at $\phi < \phi_1$ and $\phi > \phi_2$, with a smaller slope (by a factor $r < 1$) in the second region. The potential in the intermediate region is a second order polynomial chosen so that $V$ and $V'$ are continuous at both $\phi_1$ and $\phi_2$.

\begin{figure}
\centerline{
\includegraphics[width=0.5\textwidth,angle=0]{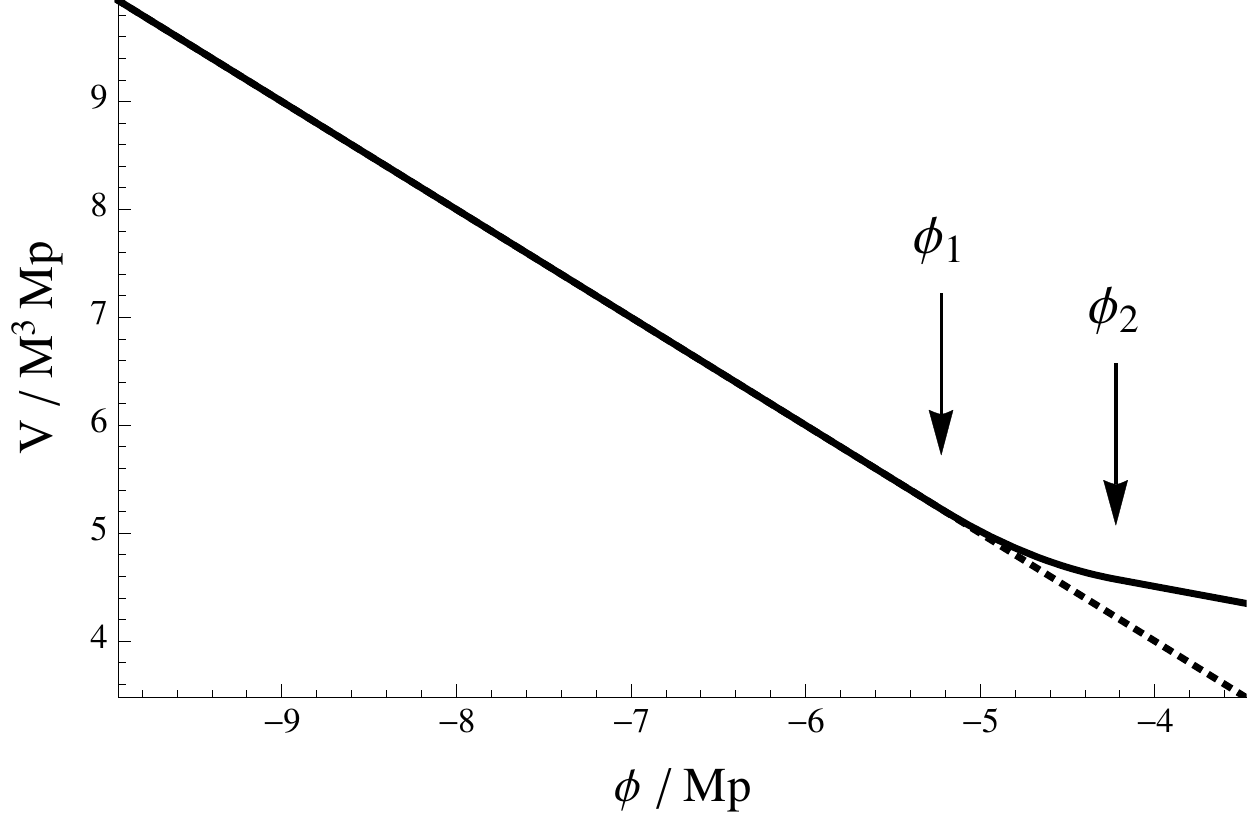}
}
\caption{The solid line shows the inflaton potential (\ref{modified-Vphi}) spanned by the inflation from $N=60$ to $N=5$, with parameters leading to the spectra of Figure \ref{fig:mod-lin-spectra}. The two arrows indicate the position of the two transition regions (the potential is linear both at $\phi < \phi_1$ and $\phi > \phi_2$, but with a different slope).  The dotted lines shows an unmodified linear inflaton potential. 
}
\label{fig:modpot}
\end{figure}

In Figure \ref{fig:modpot} we compare this modified potential (solid line) with the unmodified linear potential (dashed line). We choose parameters so that the inflaton spans $60$ e-folds of inflation in the range shown in the figure, that we assume to corresponds to an evolution between $N=60$ and $N=0$ e-folds before the end of inflation (we note that the potential (\ref{modified-Vphi}) is unbounded from below; the potential needs to be further modified at greater values than those shown, so to have a stable minimum with $V=0$). The value of $\phi_1$ is chosen so that the departure from the initial linear potential occurs at $N=24$  (this gives $\phi_1 \simeq -5.22 \, M_p$; we then choose $\phi_2 = -4.22 \, M_p$). We then choose $r =0.3$, so that the derivative of the potential decreases of a factor of about $\sim 1/3$ from $\phi_1$ to $\phi_2$.  

\begin{figure}
\centerline{
\includegraphics[width=0.35\textwidth,angle=0]{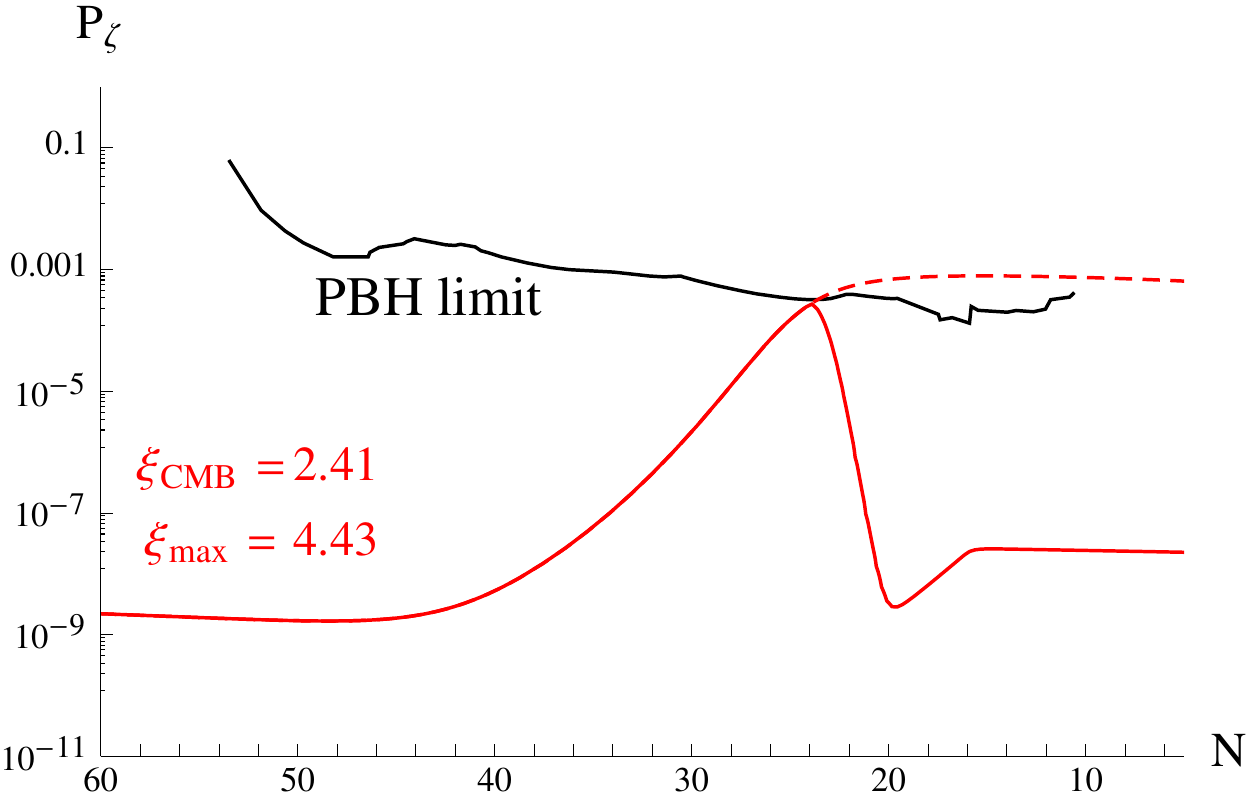}
\includegraphics[width=0.35\textwidth,angle=0]{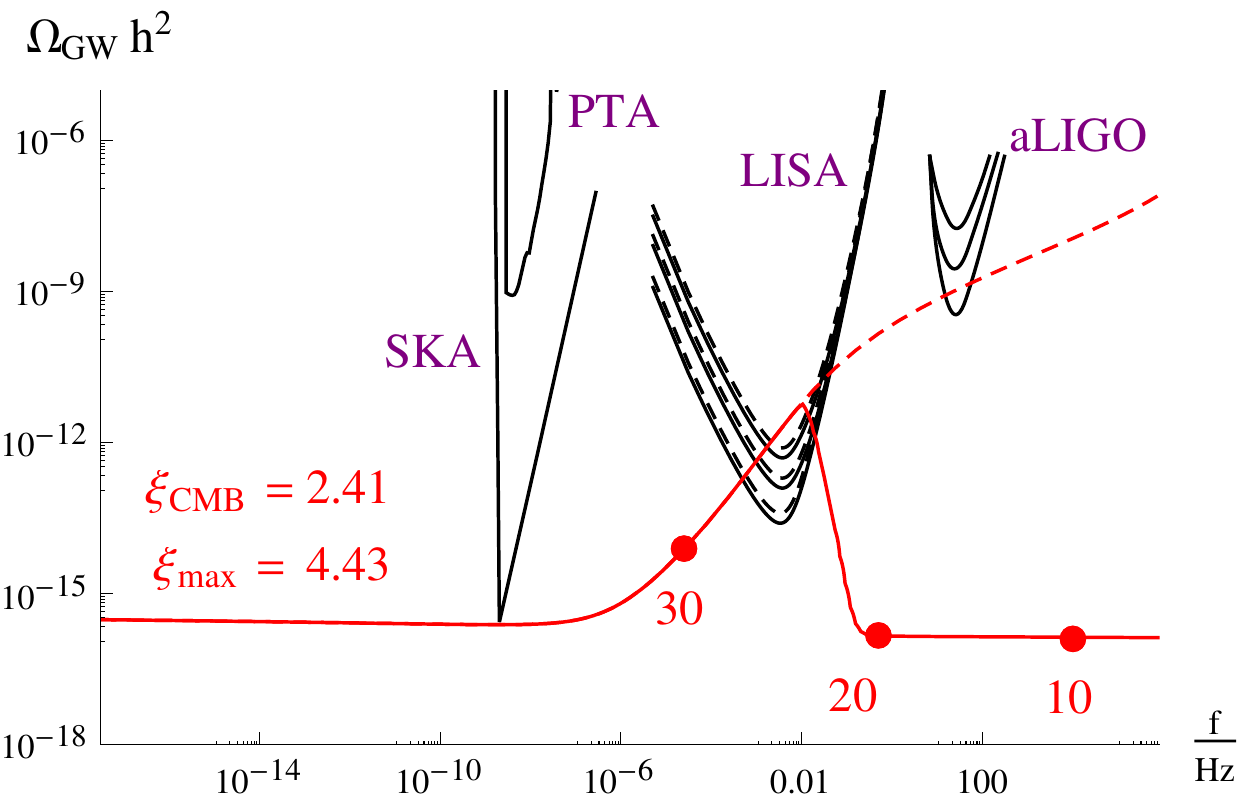}
}
\caption{As in Figure \ref{fig:lin-1p66}, but with a larger coupling of the inflaton to the gauge field, and with the modified inflaton potential (\ref{modified-Vphi}).  The solid lines are the spectra obtained in this case (the corresponding potential is shown in the solid line of Figure \ref{fig:modpot}). The dashed lines show how the spectra would continue at small scales if the instead the inflaton potential remained linear at all values (corresponding to the dashed line in Figure  \ref{fig:modpot}).  
}
\label{fig:mod-lin-spectra}
\end{figure}

We choose the inflaton-gauge field maximum coupling allowed by CMB in the case of a linear potential, $f = M_p/48$, see Ref.~\cite{Ade:2015lrj}. This corresponds to $\xi_{\rm CMB} \simeq 2.41$ at $N=60$.  The inflaton speed increases in the initial linear potential until $\phi$ reaches $\phi_1$. At this moment, $\xi \simeq 4.43$, which is within the limit of validity of perturbation theory ($\xi \la 4.8$) obtained in Ref.~\cite{Peloso:2016gqs}. The inflaton speed, and the parameter $\xi$ decrease at the transition $\phi = \phi_1$,  due to the decrease of the slope of the inflation potential. This significantly reduces the gauge field amplification and the sourced scalar and tensor modes. To quantify the effect of the change of the potential, in Figures \ref{fig:modpot} and \ref{fig:mod-lin-spectra} we also show with dashed lines the unmodified linear potential at all values of $\phi$,  and the corresponding scalar and GW spectra at small scales. We see that the relatively mild change of the potential results in very different signal for the modes that exit the horizon at those scales, due to the exponential sensitivity of the gauge field amplification to the parameter $\xi$. 

We see that the potential (\ref{modified-Vphi}) can indeed result in a visible signal at LISA scales, without violating bounds from PBH.  Many other examples can be constructed. For instance, in the next section we discuss how a localized event of gauge field amplification can be obtained in a two field model.

\subsubsection{Dependence on the number of gauge fields}  
\label{subsubsec:phi-part2B}

Let us assume that  ${\cal N} > 1$ vector fields are amplified by the ${\cal L} = - \frac{\phi}{4 f} \, F_i {\tilde F}_i$ interaction ($i=1,\, \dots ,\, {\cal N}$). For simplicity, we assume that all the fields have the same coupling to the inflaton, as for instance will happen if the vectors are the different components of a  non-abelian group. This has several consequences: (i)  an increased backcreaction, that will slow the motion of the inflaton more than in the ${\cal N} = 1$ case; (ii)  an increased GW source: as the different gauge fields are statistically uncorrelated with each other, the GW power spectrum - for any given value of $\xi$ - increases by  ${\cal N}$ with respect to the case of a single vector field;~\footnote{For any given model and coupling, this does not imply a growth of the GW power spectrum by ${\cal N}$, due to the increase backreaction on the background.} (iii) an analogous increase $\propto {\cal N}$ taking place for the power spectrum scalar perturbations, schematically, for $\zeta \propto \sum_{i=1}^{\cal N}   \vec{E}_i \cdot \vec{B}_i $, we have 
\begin{equation}
\left\langle \zeta \zeta \right\rangle \propto \sum_{i,j} \left  \langle \left( \vec{E}_i \cdot \vec{B}_i \right)  \left( \vec{E}_j \cdot \vec{B}_j \right) \right\rangle = \sum_i  \left\langle \left( \vec{E}_i \cdot \vec{B}_i \right)^2 \right\rangle = {\cal N}  \left\langle  \left( \vec{E}_1 \cdot \vec{B}_1 \right)^2 \right\rangle \,, 
 \end{equation}
 (namely, the different sources are statistically uncorrelated, resulting in an ${\cal N}$ enhancement with respect to the case of a single gauge field); this is contrasted by the fact that also the second term in (\ref{beta-lorenzo}) increases by ${\cal N}$. Therefore, as we can observe from (\ref{Pz2}), the scalar power spectrum has a ${\cal N}$ enhancement in the $\xi \ga 1$ regime, when $\beta \simeq 1$, while a $1 / {\cal N}$ suppression \cite{Anber:2009ua}  in the  $\xi \gg 1$ regime, when the second term dominates in $\beta$.  Therefore, in the $\xi \gg 1$ regime, the ratio between the GW and the scalar power spectra scales as ${\cal N}^2$. It is reasonable to expect that even mild values of ${\cal N}$ can lead to an observable GW signal, while respecting the PBH bound.

\begin{figure}
\centerline{
\includegraphics[width=0.35\textwidth,angle=0]{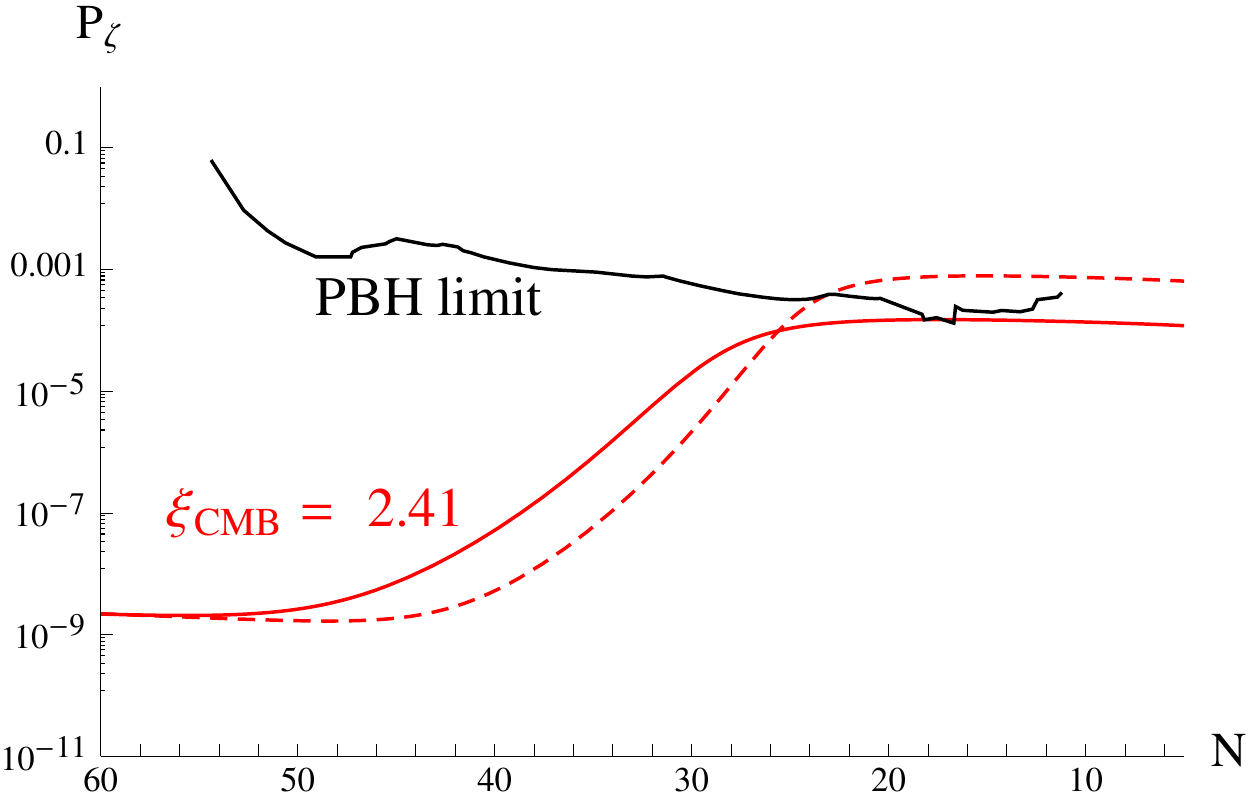}
\includegraphics[width=0.35\textwidth,angle=0]{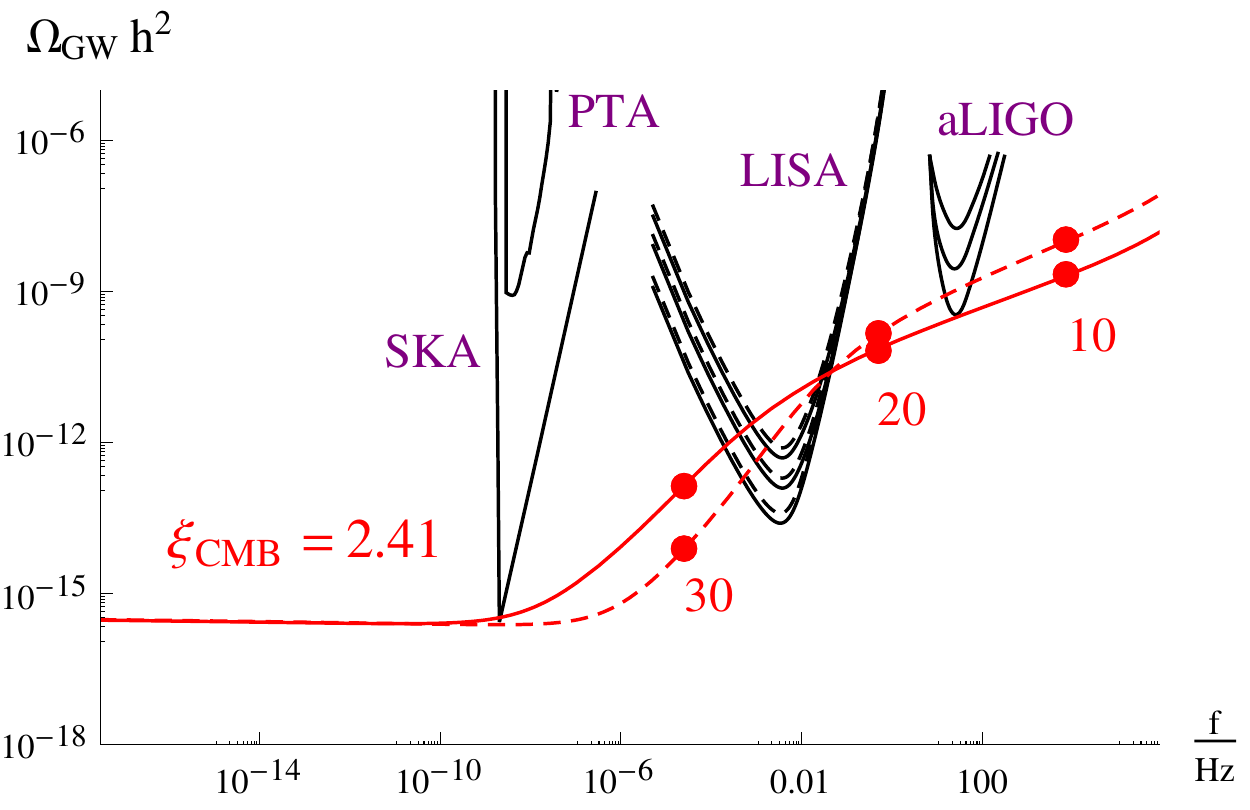}
}
\caption{Scalar and tensor signals for a linear inflation potential. The solid lines show the signal if  ${\cal N} = 6$ gauge fields are amplified. For comparison, the dashed lines show the signal when  $1$ gauge field is amplified.
}
\label{fig:N6-lin-spectra}
\end{figure}

This is confirmed by Fig.~\ref{fig:N6-lin-spectra}, where the solid (dashed) lines show the scalar and tensor power spectra generated if ${\cal N}=6$ (1) gauge fields are amplified. When comparing the solid with the dashed lines, we notice an increased GW signal at LISA scales, while the scalar signal is now below the PBH bound at all scales. We note that, for ${\cal N} = 6$, the source scalar signal is indeed enhanced at the largest $N$ shown (corresponding to $\xi \ga 1$), and suppressed at the smallest $N$ shown (corresponding to $\xi \gg 1$).~\footnote{We fixed the the coupling $f = M_p/48$ to the same value used in the previous figure, corresponding to the Planck limit for a linear potential and a single amplified gauge field ~\cite{Ade:2015lrj}. It is possible that, for ${\cal N} > 1$ a slightly smaller coupling should be considered. This would require a dedicated analysis on the CMB limits, which is beyond our scope. A slightly smaller coupling would not change our findings, and the present discussion.}  This is due in part to the transition for the $\propto {\cal N}$ enhancement to the $\propto 1/{\cal N}$ suppression that we have just discussed, and in part to the increased backreaction, that slows the inflaton, and decreases $\xi$. We see that also the GW spectrum is suppressed at small scales with respect to the case of single gauge field. This suppression is just due to the increased backreaction.

\section{Production from a rolling field $\sigma$ different from the inflaton}  
\label{sec:sigma}

In this Section we provide a different example on how to obtain a large GW at interferometer scales without conflicting with the PBH limit at $N \simeq 10$. We employ the model of  \cite{Namba:2015gja}, which provided a localized bump in the spectrum of scalar and tensor perturbations. The model was proposed to provide an explicit example that can generate a visible tensor-to-scalar ratio at the CMB scale for arbitrarily small scale of inflation (and, therefore, for arbitrary small vacuum GW), without conflicting with limits imposed by the non-gaussianity of the  scalar perturbations at CMB scales. Here we show that this model can also provide a sufficiently large GW signal for detection, and a sufficiently small scalar, at interferometer scales, particularly for the case of the PTA-SKA and LISA experiments. The model is  \cite{Namba:2015gja} 
\begin{equation}
{\cal L} = - \frac{1}{2} \left( \partial \phi \right)^2 - V_{\phi} \left( \phi \right) + \frac{1}{2} \left( \partial \sigma \right)^2  + V_{\sigma} (\sigma) + \frac{1}{4} F^2  - \alpha \frac{\sigma}{f} F {\tilde F} \;, 
\label{lagrangian-phi-sigma} 
\end{equation}
where $\phi$ is the inflaton field,  $\sigma$ is a pseudo-scalar field different from the inflaton that leads to gauge field amplification, and $\alpha$ a dimensionless parameter. For definiteness, we assume that $\sigma$ has the simplest potential typically associated to a pseudo-scalar, 
\begin{equation}
V_{\sigma} \left( \sigma \right) = \frac{ \Lambda^4}{2} \left[ \cos{\frac{\sigma}{f}}+1 \right] \;. 
\label{V-sigma}
\end{equation}
We tune the curvature of the potential to be of the same order as the Hubble rate $H$ during inflation.  We encode this tuning into the parameter 
\begin{equation}
\delta \equiv \frac{\Lambda^4}{6 H^2 f^2} \la 1 \;,  
\label{delta}
\end{equation}
so that the mass of $\sigma$ in the minimum of the potential $m_\sigma = \sqrt{3 \delta} \, H$ is slightly smaller, but comparable to $H$ (we choose $\delta = 0.2$ in our computations). The pseudoscalar then experiences the slow roll evolution  \cite{Namba:2015gja} 
\begin{equation}
\sigma =2 \, f \, {\rm arctan} \left[{\rm e}^{\delta \, H \left( t - t_* \right)} \right]  \;\;\; \Rightarrow \;\;\; \dot{\sigma} = \frac{f \, H \, \delta}{\cosh \left[ \delta \, H \left( t - t_* \right) \right] } \;.  
\label{dotsigma}
\end{equation}
We see that the evolution is non-negligible for $1/\delta \sim $ few e-folds.  In our computations below we fix $\delta =0.2$ for definiteness.  The quantity $t_*$ is the time at which $\sigma$ is in the steepest position of the potential ($\sigma = f\pi/2$), and at which it has maximum speed. This value depends on the initial conditions, and we simply treat it as a free parameter of the model.~\footnote{One could imagine a more complicated model, with a different potential $V_\sigma$ close to the origin,  that fixes the initial condition for $\sigma$ at the required position.} 

The rest of this section is divided into three parts. In the first part, we review the results on the scalar and tensor modes sourced during inflation in this model~\cite{Namba:2015gja}. In the second part, we study the prospect of detection of the inflationary GW signal at interferometers. In the third part, we discuss a different potential mechanism for GW production. Specifically, we study the possibility that the sourced scalar perturbations produce PBH (in an amount consistent with the limits shown in Figure \ref{fig:limits}), and that the merging of two such PBH in the recent universe gives rise to detectable GW by PTA-SKA, LISA, or AdvLIGO.  We stress that this is an independent mechanism of GW production with respect to the one discussed in the second part, and that the two mechanisms are actually sensitive to gauge field amplification taking place at different times during inflation. 

Before we present this analysis, we conclude this discussion by pointing out that a different way to produce a localized bump, without resorting to this second field $\sigma$,  is to assume that the vector field is massive, and that its mass is modulated by the value of the inflaton field $\phi$.~\footnote{We thank Andrei Linde for suggesting this possibility.} Refs.  \cite{Meerburg:2012id,Linde:2012bt} also studied the possibility that the gauge field has a mass $> \xi H$, which highly suppresses the gauge field amplification, and the consequent PBH production. If the mass depends on the inflaton, in such a way that the vector is light only in a neighborhood of some given value $\phi = \phi_*$ assumed during inflation, then only the  gauge modes produced while $\phi$ spans this interval are produced, generating an enhanced scalar and tensor signal at these scales, similarly to the one that we consider in this section.

\subsection{Production of scalar and tensor modes }
\label{subsec:phisigma-part1}

For sufficiently large coupling strength $\alpha/f$, the motion of $\sigma$ can lead to a strong amplification of the vector modes that leave the horizon during the $\sim 1/\delta$ e-folds of inflation in which $\sigma$ has a non-negligible roll~\cite{Namba:2015gja}.  In turn,  the amplified gauge modes source perturbations of  $\sigma$ and gravity waves, through the $2 \rightarrow 1$ processes that we have already discussed  in the previous section. During this time, the inflaton field is also rolling; the simultaneous roll of $\phi$ and $\sigma$ gives rise to a $\delta \sigma - \delta \phi$ coupling between the perturbations of the two fields.  As long as this coupling is active, the perturbations $\delta \sigma$ produced by the vector field source inflaton perturbations,\footnote{This channel of production of inflaton perturbations is more efficient \cite{Ferreira:2014zia}  than the gravitational production of inflaton perturbations from the gauge modes \cite{Barnaby:2012xt}.} which give rise to a bump in the primordial scalar curvature.\footnote{  The primordial scalar curvature $\zeta$ is a linear combination $\zeta \left( t \right) = A \left( t \right) \delta \phi \left( t \right) + B \left( t \right) \delta \sigma  \left( t \right)$, where the two coefficients $A$ and $B$ depend on the background. In this model, the field $\sigma$  becomes massive shortly after $t_*$.  As $\sigma$ becomes a massive field in an inflationary universe, its energy density and pressure rapidly drops to zero, and so does the coefficient $B$. The only potentially observable effect of $\delta \sigma$ is through its linear coupling to the inflaton perturbations that we are considering, and that is in act only as long as $\dot{\sigma} \neq 0$.}  

The bumps in the primordial scalar and tensor perturbations due to this mechanism add us incoherently to the standard modes from the vacuum, and the total power spectra are given by  \cite{Namba:2015gja} 
\begin{eqnarray}
P_{\zeta} \left( k \right) &\simeq&   P_{\zeta,v} \left( k \right) +  \left[ \frac{H^2 \left( k \right)}{8 \pi^2 \, M_p^2}  \right]^2 \,  f_{2,\zeta} \left( \frac{k}{k_*} ,\, \xi_* ,\, \delta  \right)  \;\;, \nonumber\\ 
P_{\rm GW} \left( k \right) &\simeq&  16 \, \epsilon_\phi   P_{\zeta,v} \left( k \right)  +  \left[  \frac{H^2 \left( k \right)}{8 \pi^2 \, M_p^2}   \right]^2 \,  f_{2,+} \left( \frac{k}{k_*} ,\, \xi_* ,\, \delta  \right)  \;\;,  
\label{Pz-PGW-model2}
\end{eqnarray} 
where again we disregard the sourced $-$ GW polarization, which is much smaller than the  sourced $+$ polarization. In this expression we have denoted the inflaton slow-roll parameter with $\epsilon_\phi$, to distinguish it from the slow roll parameter $\epsilon_\sigma \equiv \dot\sigma^2/2 H^2 M_p^2$ associated with the field $\sigma$. The quantity $k_*$ denotes the comoving momentum of the mode that leaves the horizon at $t_*$, while $\xi_* \equiv \alpha \, \dot\sigma_*/2 f H$ with $\dot{\sigma}_* \equiv \sigma \left( t_* \right)$. 

Ref.  \cite{Namba:2015gja} evaluated the sourced power spectra numerically for the two specific examples $\delta = 0.2,0.5$ and for several values of $\xi_*$, and showed that they are well fitted by 
\begin{equation}
f_{2,j}  \left( \frac{k}{k_*} ,\, \xi_* ,\, \delta \right) \simeq  f_{2,j}^c \left[ \xi_* ,\, \delta \right] \, {\rm exp} \left[ - \frac{1}{2 \sigma_{2,j}^2 \left[ \xi_* ,\, \delta \right]} \, {\rm ln }^2 \left( \frac{k}{k_* \; x_{2,j}^c \left[ \xi_* ,\, \delta \right] } \right) \right] \;\;\;,\;\;\; j = \zeta ,\, + \;\;, 
\label{f2-j}
\end{equation} 
which is a Gaussian bump (in terms of $\ln k$) centered at $k = x_{2,j}^c \, k_*$. The quantity $x_{2,j}^c$ is of  $\calO$ so that, as expected, the sourced signals are peaked at the scales that leave the horizon close to the time at which the roll of $\sigma$ is fastest.  The function $f_{2,j}^c $ controls the amplitude of the bump, and, analogously to (\ref{Pz-PGW}), it grows exponentially with $\xi_*$.  The function $ \sigma_{2,j}^2 $ controls the width of the bump, and it decreases with increasing $\delta$. This is also to be expected, since greater $\delta$ corresponds to a shorter duration of the roll of $\sigma$. The precise dependence of these three functions on $\xi_*$ is given in Ref.~\cite{Namba:2015gja}, for the two cases  $\delta =0.2,\, 0.5$.

\subsection{ Direct Detection of Inflationary Signatures at Interferometer Scales}
\label{dirdet}

To obtain the precise scalar and tensor perturbations generated in the model (\ref{lagrangian-phi-sigma}) we need  to specify the inflaton potential. From the potential we derive  the slow roll parameters $\epsilon_\phi \equiv \frac{M_p^2}{2} \left( \frac{\partial_\phi V_\phi}{V} \right)^2$ and  $\eta_\phi \equiv M_p^2  \frac{\partial_{\phi\phi} V_\phi}{V}$. As long as the background and the perturbation contributions from $\sigma$ and the vector fields can be neglected, we recover the standard results $\dot{H} \simeq - \epsilon_\phi \, H^2$ for the evolution of the Hubble rate, $n_s \simeq 1 + 2 \eta_\phi - 6 \epsilon_\phi$ for the tilt of the scalar spectrum, and $r \simeq 16 \epsilon_\phi$ for the tensor-to-scalar ratio.  For definiteness, we assume that $r=0.01$ (parametrically close to the current bound   $0.07$ at $95\%$ confidence \cite{Array:2015xqh}), giving $\epsilon_\phi \simeq 6.25 \cdot 10^{-4}$. From the observed value $n_s -1 \simeq -0.035$ \cite{Ade:2015lrj} we then obtain $\eta_\phi \simeq - 0.015$. Having $\vert \eta_\phi \vert \gg \epsilon_\phi$ is for example typical of top-hill inflationary potentials~\cite{Garcia-Bellido:2014gna}. 

In the examples that we show in this subsection we avoid specifying an inflationary potential, and we assume that $\epsilon_\phi$ and $\eta_\phi$ remain constant all throughout inflation. It is immediate to modify this, once a  specific $V_\phi$ is given. However, since the slow roll parameters  vary at second order in slow roll, this approximation is sufficiently adequate for our purposes, and it does not affect our general conclusions for the mechanism that we are studying in this section. Therefore, we take 
\begin{equation}
H \left( N \right) = H_{\rm CMB} \, {\rm e}^{- \epsilon_\phi \left( N_{\rm CMB} - N \right)} \;\;,\;\; 
P_{\zeta,v} \left( k_N \right) = P_{\zeta,v} \left( k_{N_{\rm CMB}} \right) \, {\rm e}^{- \left( 1 - \epsilon_\phi \right) \left( 1 - n_s \right) \left( N_{\rm CMB} - N \right)} \;\;. 
\end{equation} 
(where, clearly, $\epsilon_\phi$ can be disregarded in the second relation), with $\epsilon_\phi$, $n_s$, fixed at the CMB scales. We then assume $N_{\rm CMB} = 60$, and take $P_\zeta \left( k_{N_{\rm CMB}} \right) \simeq 2.2 \cdot 10^{-9}$  \cite{Ade:2015lrj}.

\begin{figure}[ht!]
\centerline{
\includegraphics[width=0.5\textwidth,angle=0]{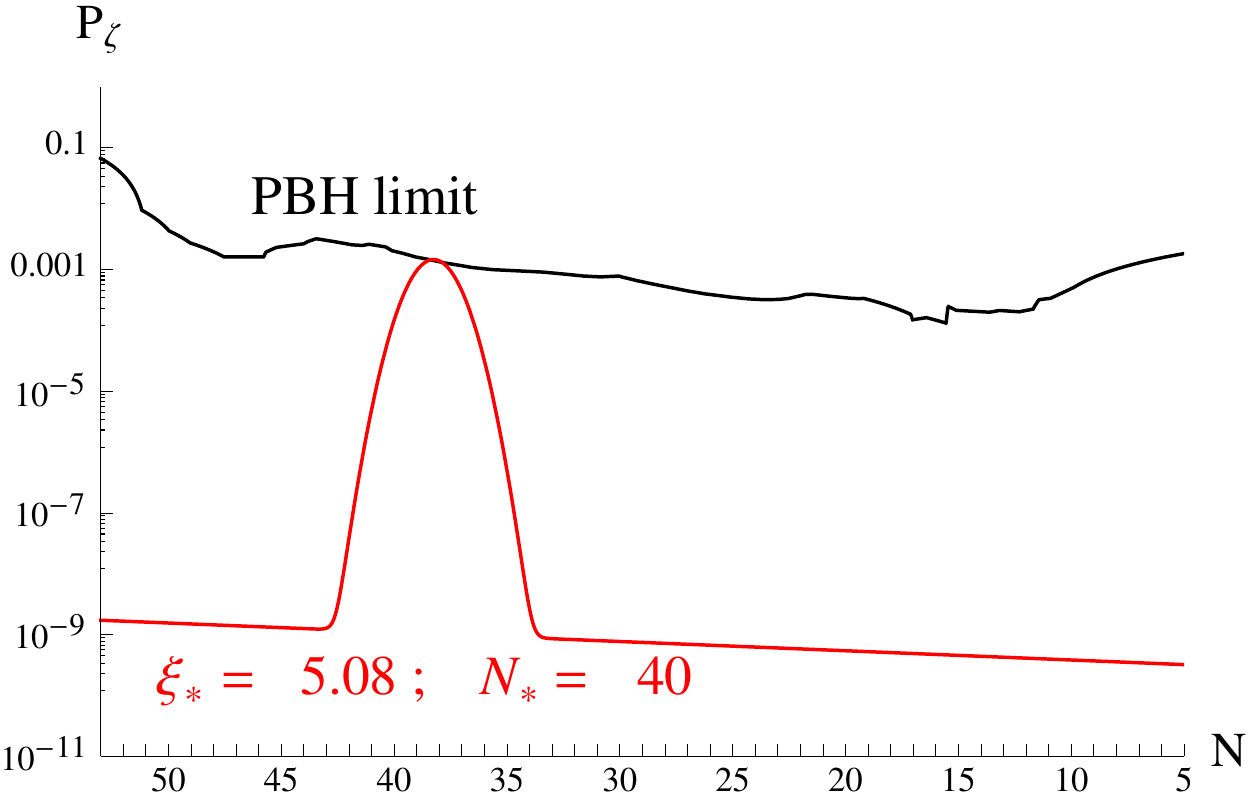}
\includegraphics[width=0.5\textwidth,angle=0]{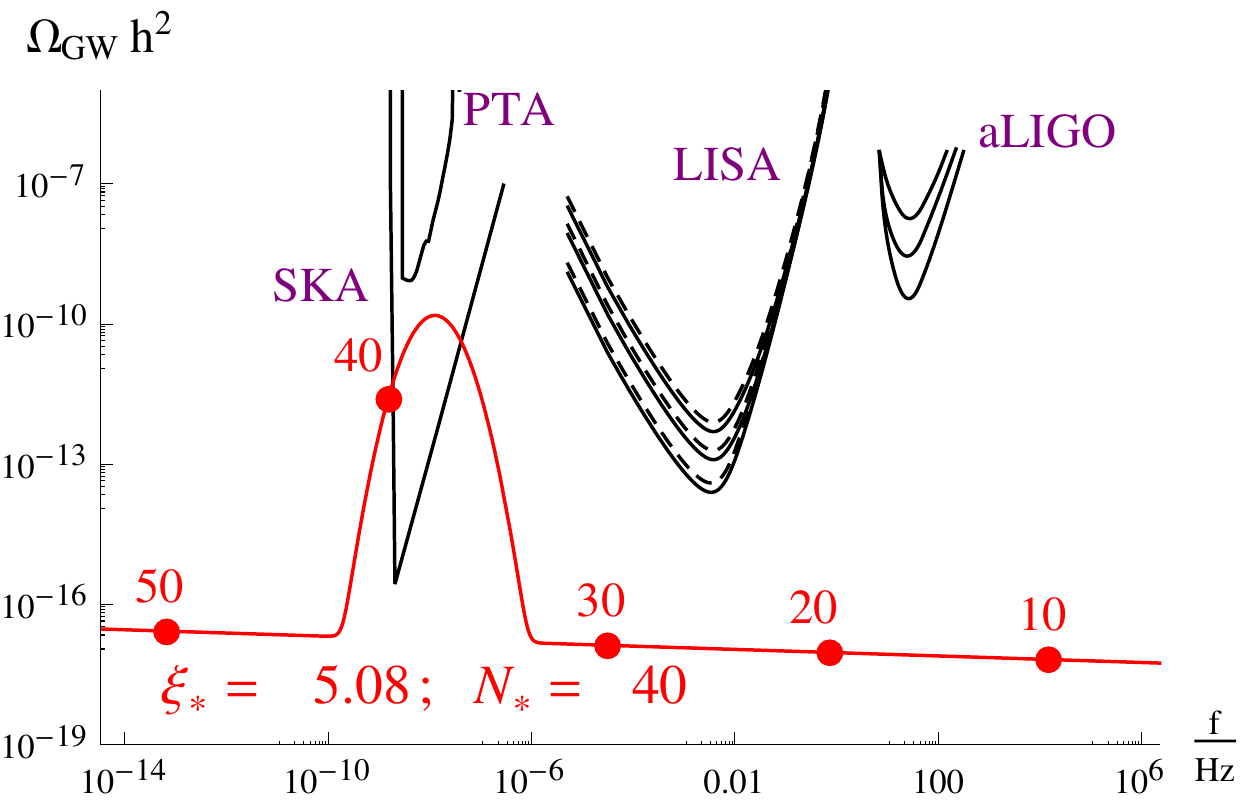}
}
\caption{Scalar power spectrum and GW power spectrum produced during inflation in the two field mode (\ref{lagrangian-phi-sigma}), assuming a bump at PTA scales. We see that, it is possible to produce a visible GW signal without violating the PBH bounds.
}
\label{fig:PTA-bump}
\end{figure}

\begin{figure}[ht!]
\centerline{
\includegraphics[width=0.5\textwidth,angle=0]{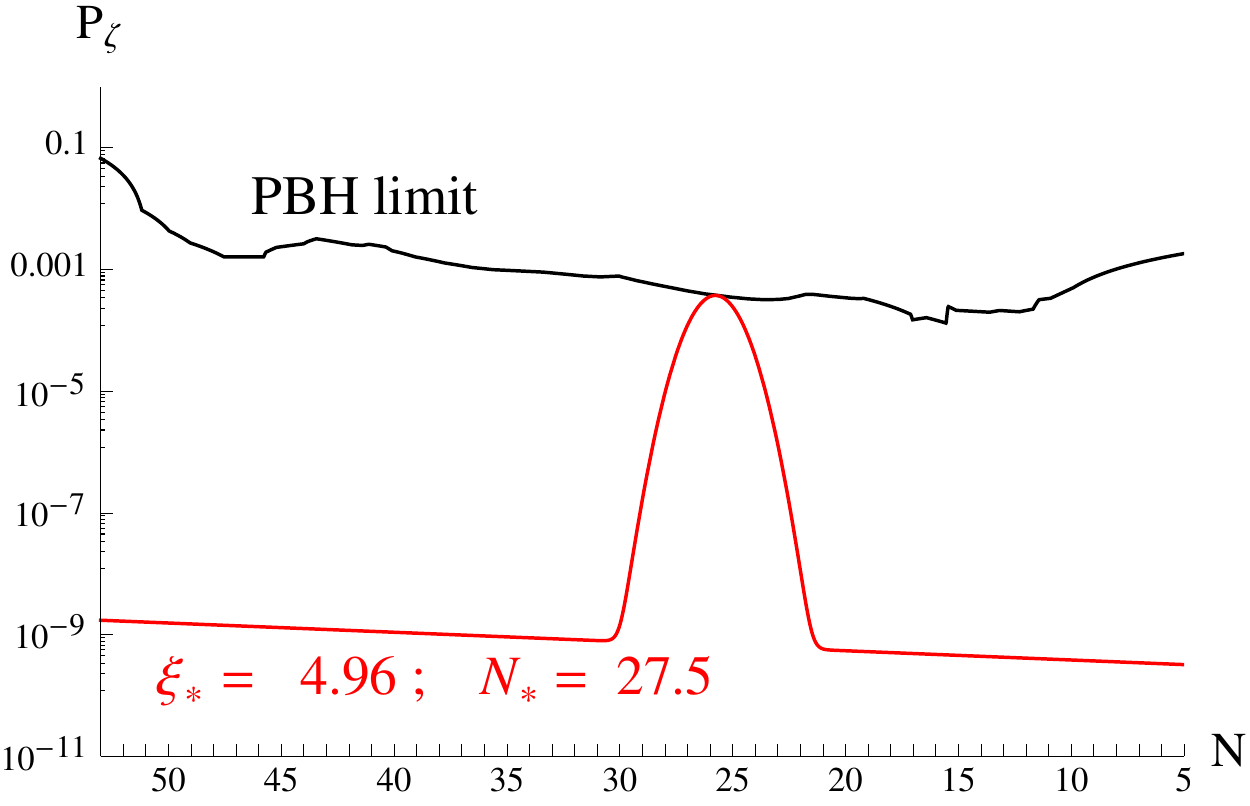}
\includegraphics[width=0.5\textwidth,angle=0]{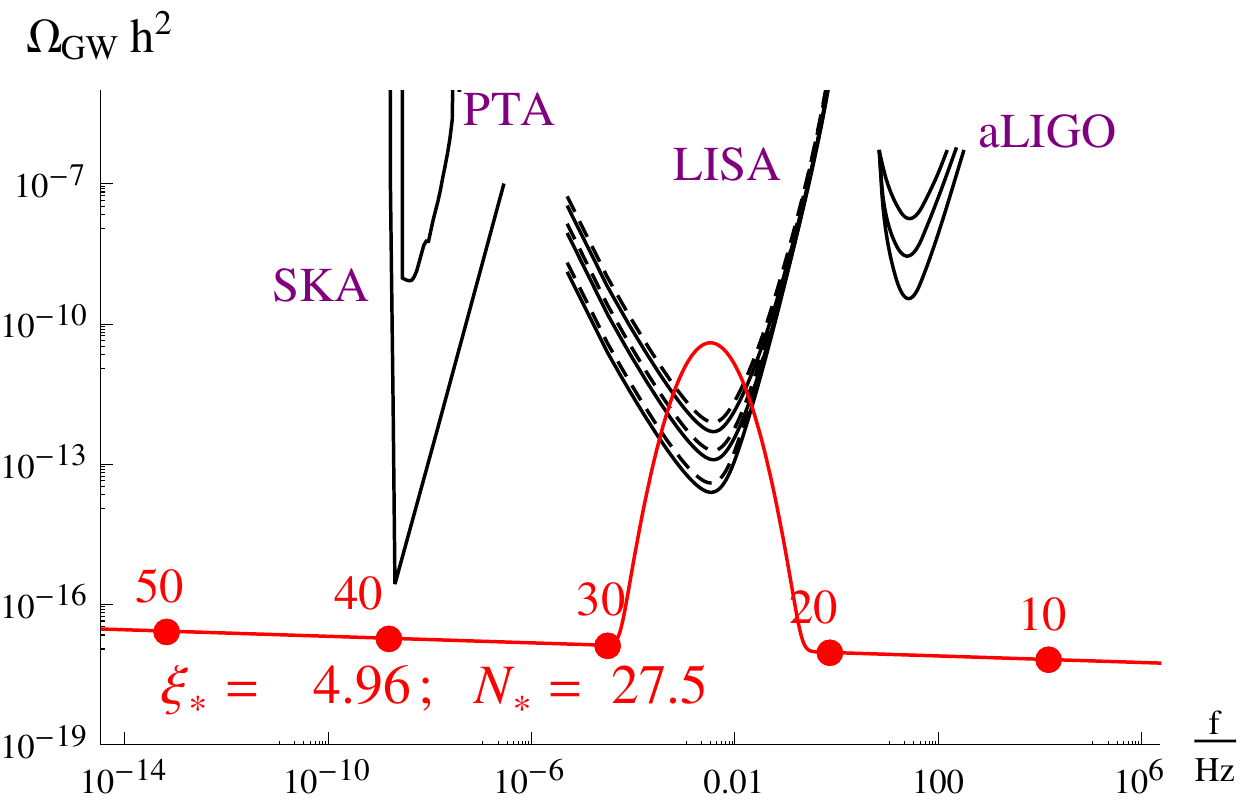}
}
\caption{Scalar power spectrum and GW power spectrum produced during inflation in the two field mode (\ref{lagrangian-phi-sigma}), assuming a bump at LISA scales. We see that,  similarly to the example shown in Figure \ref{fig:mod-lin-spectra}, it is possible to produce a visible GW signal without violating the PBH bounds.
}
\label{fig:LISA-bump}
\end{figure}

With these assumptions, the mechanism only depends on the three parameters $\xi_* ,\, N_* $ and $\delta$ discussed in the previous subsection, that control the dynamics of $\sigma$ and the field amplification. More precisely, $N_*$ is the number of e-folds corresponding to $t_*$, at which the motion of $\sigma$ is fastest. For definiteness, we take $\delta = 0.2$, corresponding to a roll of sigma for $\Delta N \simeq 1/\delta = 5$ e-folds, and to a comparable width of the sourced signal. We then choose $N_*$ so that the peak of the sourced GW signal is either at PTA, LISA or AdvLIGO scales, and we choose $\xi_*$ so that the sourced scalar modes saturate the PBH bounds. We then study whether this value of $\xi_*$ is enough to provide a visible GW signal.

\begin{figure}[ht!]
\centerline{
\includegraphics[width=0.5\textwidth,angle=0]{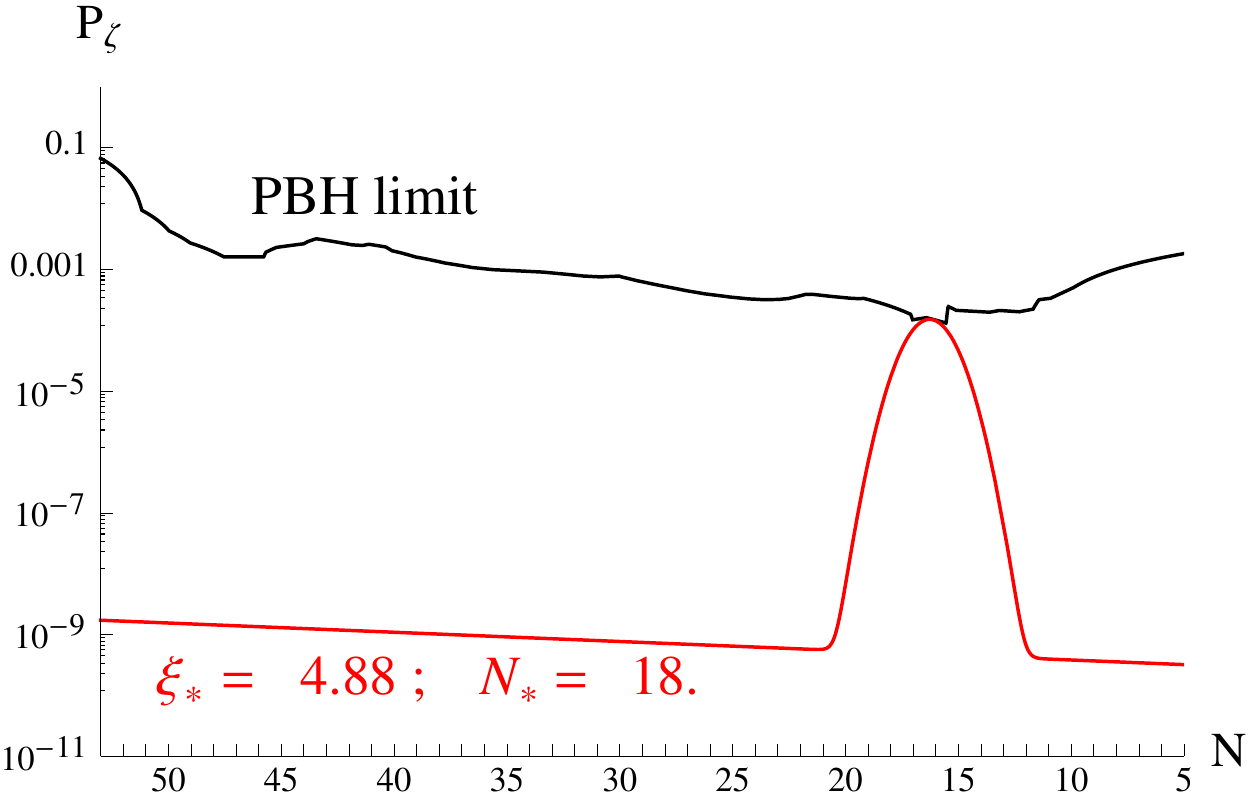}
\includegraphics[width=0.5\textwidth,angle=0]{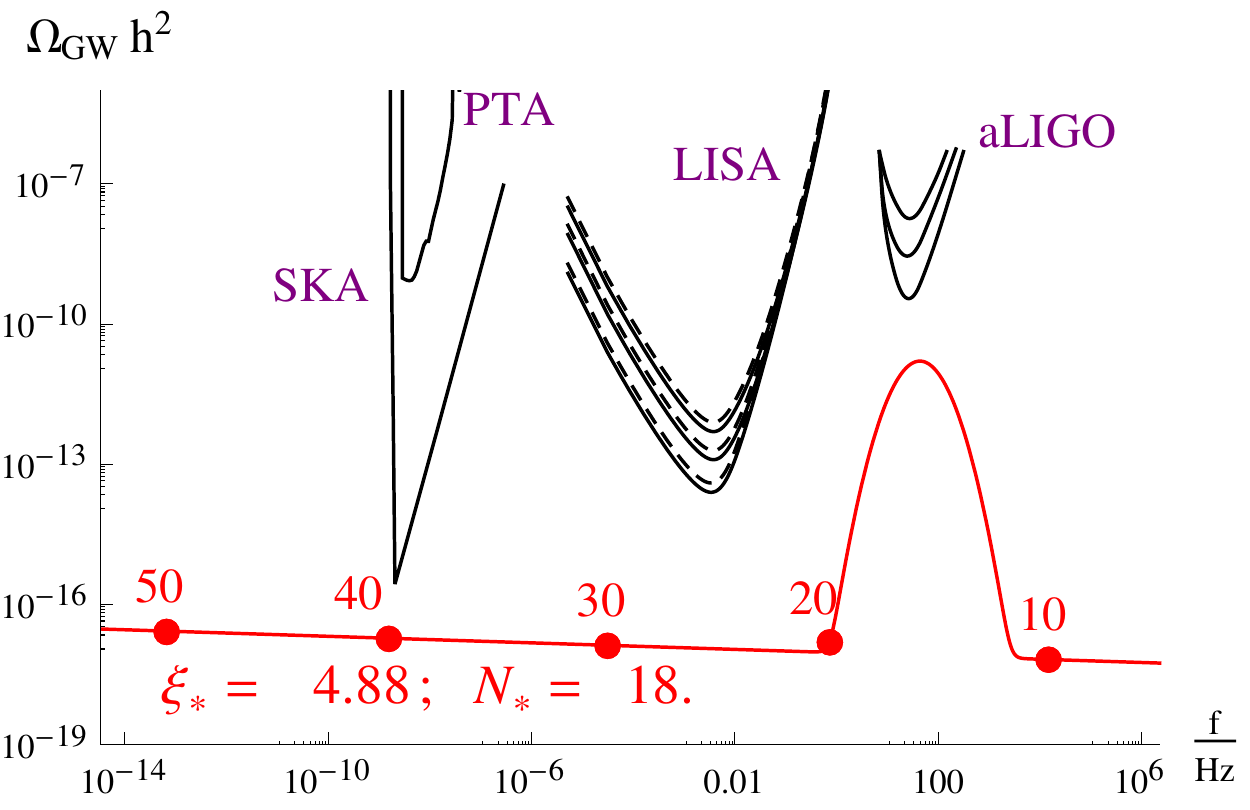}
}
\caption{Scalar power spectrum and GW power spectrum produced during inflation in the two field model (\ref{lagrangian-phi-sigma}), assuming a bump at AdvLIGO scales. 
}
\label{fig:aligo-bump}
\end{figure}

The results of  Figures \ref{fig:PTA-bump} and \ref{fig:LISA-bump} show that it is indeed possible to obtain a visible signature at, respectively, PTA and LISA scales. The same does not appear to be true for a signal at the AdvLIGO scale, and with the  AdvLIGO sensitivity, as can be seen from Figure \ref{fig:aligo-bump}. Analogous conclusions can be reached from the single field model (\ref{lagrangian-phi}) of Section \ref{sec:phi}, as can be seen most immediately from the example shown in Figure \ref{fig:mod-lin-spectra}. From the comparison of the two models, we believe that this is a generic feature associated to this mechanism. We stress that the negative concliusion reached on the AdvLIGO case depend on the computed scalar spectrum,  which is more uncertain than the GW one.

The computations performed in this section have been obtained from the  two-field model (\ref{lagrangian-phi-sigma}), under the assumptions that   (i) the gauge field amplification backreacts in a negligible way on the  background solution and (ii) the perturbations remain in the perturbative regime. Ref.  \cite{Peloso:2016gqs} showed that this is the case provided the axion scale $f$ is within a certain interval, given by the expression  (6.3) of Ref.  \cite{Peloso:2016gqs}. Combining eqs. (3.4) and (3.5) of  \cite{Peloso:2016gqs}, one obtains $f_{+,2}^c \left[ \xi_* ,\, \delta = 0.2 \right] \simeq 3.6 \cdot 10^{-5} \, {\rm e}^{3.48 \pi \xi_*}$. Using eqs. (\ref{OmGW}) and (\ref{Pz-PGW-model2}), this relation gives 
\begin{equation}
\left( \Omega_{\rm GW} \, h^2 \right)_{\rm peak}^{1/4} \simeq 1.32 \cdot 10^{-7} \, \sqrt{\epsilon_\phi} \, {\rm e}^{2.74 \, \xi_*}  \;,  
\end{equation}  
at the peak of the sourced GW signal  (we are disregarding the variation of $H$ during inflation, which provides a negligible correction to this relation for the values $r_{\rm vacuum} = 0.01$ considered here). Using this relation, the condition  (6.3) of Ref.  \cite{Peloso:2016gqs} rewrites 
\begin{equation}
\left( \frac{\Omega_{\rm GW} \, h^2}{2 \cdot 10^{-9} } \right)_{\rm peak}^{1/4}   \la \frac{f}{M_p} \la 1 \;. 
\label{condition} 
\end{equation} 
The peak values of the GW signals shown in  Figures \ref{fig:PTA-bump},  \ref{fig:LISA-bump}, and  \ref{fig:aligo-bump} is always below 
$2 \cdot 10^{-9} $. Therefore, the condition (\ref{condition}) is satisfied in a nonvanishing interval for $f$. The interval grows at decreasing values of the peaked GW signal.

\subsection{GW from merging PBHs } 
\label{subsec:collision}

In this subsection we discuss the GW emission of inspiraling primordial BHs formed when the enhanced curvature fluctuations re-enter the horizon during the radiation epoch. Such PBH present a broad mass distribution and satisfy the bounds in Figure \ref{fig:limits} while still comprising all of the Dark Matter~\cite{Clesse:2015wea}. When they form, their contribution to the energy density of the universe is negligible and do not affect BBN, but by the time of matter-radiation equality they begin to dominate the total energy density. After recombination, these PBH start to merge through hierarchical structure formation and acquire today a mass distribution peaked around a hundred solar masses. These PBHs would act as seeds for structure formation~\cite{Clesse:2015wea}. Fewer heavier PBH may have acquired significantly more mass today due to gas accretion, as well as merging, and constitute the supermassive BHs at the centers of galaxies and quasars.\footnote{For a recent analysis of the growth of an initial black hole seed up to supermassive BHs, see Ref.~\cite{Sugimura:2016dmy}.} 

The exact numbers and mass distributions are still uncertain and require detailed N-body simulations to compute the transfer function from the primordial mass distribution to the present PBH distribution. In particular, merging between different PBH will create PBH of larger masses, shifting the mass distribution. We will roughly estimate a factor $10^3$ increase in mass due to the ordinary growth of fluctuations during the matter era for the PBHs at the peak of the distribution~\cite{Clesse:2016vqa}, and at least a factor $10^5$ for the extra increase due to gravitational collapse of gas onto the high end (more massive) part of the distribution.\footnote{There will also be a decrease in the amplitude of the whole mass distribution due to the energy loss in gravitational waves from the merging of PBH, but we will ignore it here.} The first population will be responsible for the emission of GW in the AdvLIGO band and the enhanced tail of the distribution will correspond to the IMBH and SMBH population, responsible for the GW emission in the PTA and LISA bands. Note that the broad peak in the spectrum of fluctuations is responsible also for a clustering of PBHs upon reentry, which explains the rapid growth in mass and the enhanced rate of events~\cite{Clesse:2016vqa} with respect to the stellar black holes of similar mass. Keeping this growth of mass into account, we find that, as discussed in the Introduction, AdvLIGO, LISA and PTA are sensitive, respectively, to modes that left the horizon about $N \sim 35 - 37$, $N \sim 38 - 41$, and $N \sim 41-44$  e-folds before the end of inflation. 

We assume that an episode of localized field amplification, as the one obtained from the model (\ref{lagrangian-phi-sigma}) generates a bump in the scalar perturbations (as the one given by Eqs.~(\ref{Pz-PGW-model2}) and (\ref{f2-j})) in one of these three intervals. We then assume that this  initial bump produced PBH that acts as a seeds of a population of current black holes that can be identified with the dark matter of the universe~\cite{Clesse:2016vqa}. Sizable PBH in the model  (\ref{lagrangian-phi-sigma}) seeds can be generated for $\xi_* \sim 4.5 - 5$, which is compatible with limits from perturbativity~\cite{Peloso:2016gqs}.

We are interested in the collision rate of the PBH in these distributions. For a generic PBH mass spectrum, the number of collisions per volume and time is given by 
\begin{equation}
\bar \Gamma^{\rm tot}\equiv  \frac{1}{2} \sum_{A,B}  \left\langle n_{\rm BH} \left( M_A,\, t_0 \right) \,n_{\rm BH} \left( M_B,\, t_0 \right)  \, \sigma(M_A, \, M_B,\, v) \, v \right\rangle \,, 
\label{genrate}
\end{equation} 
where $n_{\rm BH} \left( M,\, t_0 \right) $ is the number density of black holes with mass $M$ at the present time $t_0$, and where  $ \sigma(M_A, \, M_B,\, v) $ is the cross section (merger rate) for the merging (inspiraling) of one black hole of mass $M_A$ on one black hole of mass $M_B$, colliding at the relative velocity~$v$. Finally, $\langle \dots \rangle$ denotes average over volume and velocity. 
 
To perform our estimate, we assume a very narrow distribution of black hole masses corresponding to a relatively large value of $\delta$ in the model  (\ref{lagrangian-phi-sigma}); in this case, we can assume a single value for the mass in the cross section  (\ref{genrate}), see~\cite{Mouri:2002mc,Clesse:2016vqa},
\begin{equation}
\sigma \left( M_{\rm BH} ,\, M_{\rm BH},\, v \right) = 2\pi \,  \left( \frac{85\,  \pi \,  2^5}{6 \sqrt{2}}   \right)^{2/7}\, \frac{G^2 \, M_{\rm BH}^2} {c^4}    \left( \frac{c}{v} \right)^{18/7} \simeq 10^{-37} \, {\rm Mpc}^{2} \, \left( \frac{M_{\rm BH}}{M_\odot} \right)^2 \left( \frac{c}{v} \right)^{18/7} \;, 
\label{crosssec}
\end{equation}
(where $G$ is Newton's constant, and $c$ the speed of light). As we are assuming that the black holes are the cold dark matter of the universe, the local  black hole number density at any place in the universe is just the local dark matter number density:  
\begin{equation}
n_{\rm BH} \left( t_0 \right) = \frac{\delta_{\rm local}}{M_{\rm BH}} \, \Omega_{\rm CDM} \, \rho_c \left( t_0 \right) \simeq 
3.3 \cdot 10^{10} \, {\rm Mpc}^{-3} \, \frac{\delta_{\rm local}}{M_{\rm BH} / M_\odot} \;, 
\end{equation}
where $\Omega_{\rm CDM} h^2 = 0.12$ has been used \cite{Ade:2015xua}, and where $\delta_{\rm local} \equiv  \rho_{\rm CDM}^{\rm local} /\rho_{\rm CDM}^{\rm mean} $ is the local enhancement of PBH dark matter in compact halos. Inserting all this in (\ref{genrate}), gives 
\begin{equation}
\bar \Gamma^{\rm tot} \simeq 2 \cdot 10^{-7} \, {\rm Gpc}^{-3} \, {\rm yr}^{-1} \, \delta_{\rm local} \, \left( \frac{10 \, {\rm km} / {\rm s}}{v} \right)^{11/7} \;,  
\end{equation} 
where we have normalized the velocity to typical velocities of compact masses in dwarf spheroidal galaxies, which are indeed in the few km/s range, up to 15 km/s~\cite{Drlica-Wagner:2015ufc}. We see that a reasonable local enhancement  in the number density could make the signal observable. For example, this is the case of GWs from inspiralling of clustered PBHs in dense compact halos building up and orbiting around galaxies~\cite{Bird:2016dcv,Clesse:2016vqa}. Some of these have been recently discovered by DES around the Milky Way and show huge mass-to-light ratios of 500-1000~\cite{Drlica-Wagner:2015ufc}. 

The possibility that PBHs comprise all of the Dark Matter is a very attractive scenario~\cite{GarciaBellido:1996qt,Clesse:2015wea} that has received special attention lately~\cite{Bird:2016dcv,Clesse:2016vqa,Sasaki:2016jop}, since the detection by AdvLIGO of GW from the merging of very massive black hole binaries~\cite{Abbott:2016blz,Abbott:2016nmj}. The systematic detection, by terrestrial laser interferometers, of merging BH binaries within a broad range of masses, will open a new window into the early universe. The characterization of the mass spectrum will tell us about the time and duration during inflation of new phenomena like particle production or new couplings of the inflaton to other scalar fields.

\section{Summary}
\label{sec:conclusions}

In this paper we studied the possibility of detecting inflationary GW in terrestrial (such as AdvLIGO) and spacial (LISA) interferometers. We specifically studied GW produced by gauge fields, amplified due to their coupling $\phi \,F {\tilde F}$ to a pseudo-scalar inflaton $\phi$, or to their coupling $\sigma F {\tilde F}$ to a different field $\sigma$ that is rolling for a few e-folds during inflation. The sourced GW are exponentially sensitive to the speed of the inflaton, so this mechanism is naturally enhanced at late times / small scales \cite{Cook:2011hg,Barnaby:2011qe}, possibly opening new windows of exploration for inflation model building. Like in the case of the detection of sourced GW at CMB scales~\cite{Barnaby:2010vf,Barnaby:2012xt,Ferreira:2014zia,Mirbabayi:2014jqa,Namba:2015gja}, the main difficulty in producing a large sourced GW signal at interferometer scales is the unavoidable simultaneous production of scalar perturbations. In this case the problem is the possible overproduction of PBH from the enhanced scalar signal~\cite{Linde:2012bt}. 

In this work we studied whether and under which conditions the limit associated with the scalar production can be circumvented. An important consideration is that there is an intrinsic uncertainty associated to the computation if the scalar perturbations for the couplings that are necessary to produce a visible GW signal. If the correct result is $\calO$-suppressed with respect to the current estimates, then the PBH limit may not be a problem~\cite{Linde:2012bt}. Ultimately, we believe that only a numerical computation of the scalar perturbations, along the lines of Refs.~\cite{Adshead:2015pva,Cheng:2015oqa,Adshead:2016iae,Notari:2016npn}, will resolve this issue. Unless proven incorrect, let us  assume that the present estimates are sufficiently accurate. In this case, we note that, given the typical blue nature of the sourced signal, the PBH limit is typically enforced by the smallest scales at which it exists, namely for modes that left the horizon around $10$ e-folds before the end of inflation. We therefore studied the constraints imposed by the PBH limits at  a given interferometer scale, without making assumptions on the later evolution of the inflaton (in essence, on the inflaton potential at different field values from those experimentally probed).  We found that the projected LISA sensitivity \cite{Bartolo:2016ami} is good enough to allow for the detection of a GW signal, in a regime that evades the PBH bounds. We instead obtained a negative conclusion  for the expected AdvLIGO sensitivity~\cite{TheLIGOScientific:2016wyq}.~\footnote{We do not have a no-go theorem in support of this statement, but only the evidence given by the two examples that we have studied. In particular, the two field $\phi-\sigma$ model is ``designed'' \cite{Namba:2015gja} to maximize the ratio between tensor and scalar perturbations, since it avoids a direct coupling between the inflation and the gauge fields, therefore constituting an optimized situation. The example we studied, and that we showed in Figure \ref{fig:aligo-bump}, led to a GW signal that is about $20$ times smaller than the best projected AdvLIGO sensitivity.} 

A different possibility to circumvent the PBH limit is to assume that ${\cal N}$ gauge fields are amplified by the same mechanism. In this case, the ratio between the tensor and the scalar power spectra is enhanced by ${\cal N}^2$ in the regime of strong coupling required to produce an observable GW signal. We confirmed the finding of \cite{Domcke:2016bkh}, that studied this possibility in the context of Starobinsky inflation \cite{Starobinsky:1980te}, showing that in the case of chaotic inflation even the moderate value ${\cal N} = 6$ allows for a visible signal at LISA and AdvLIGO, while respecting the PBH bounds.

While in most of this work we have studied the possible detection of the stochastic GW signal directly produced during inflation, in Section \ref{subsec:collision} we have discussed the alternative possibility that  density perturbations are produced during inflation collapse to form PBH, which then evolve to the present universe, and ultimately give rise to BH-BH binary mergers, such as those observed by the AdvLIGO detectors. The possibility that these BH can be identified with the dominant component of the dark matter of the universe  has been the object of interesting recent works~\cite{Clesse:2015wea,Bird:2016dcv,Clesse:2016vqa,Sasaki:2016jop,Kashlinsky:2016sdv,Kawasaki:2016pql}. The generation of these PBH requires enhanced density perturbations with respect to the amplitude at the CMB scale. In the literature, a broad peak in the density power spectrum, and the consequent PBH production has typically been obtained in the context of hybrid inflation, starting from the work \cite{GarciaBellido:1996qt}. Here, we have discussed an alternative mechanism for the generation of this broad peak, in the context of sourced perturbations in axion inflation \cite{Linde:2012bt}. The main difference between the two cases, is that the sourced perturbations are highly non-gaussian, which, at any fixed amplitude of the two-point function, results in an increased PBH fraction (see Figure \ref{fig:Pz-beta}). 

In Table \ref{tab:window} we have listed the (approximate) interval of modes that can potentially give an observable GW signatures in the various experiments. We note the presence of an interesting correspondence between modes listed in the 6th row, with the modes listed in the 4th row of the table (and, marginally, with also those in the 5th row). This opens the interesting possibility that the same event of localized particle production during inflation can give rise to both a stochastic GW signal at PTA scales, and to scalar perturbations that eventually result in BH collisions in the present universe and that are observed at LISA (and, possibly, also at AdvLIGO). The PTA signal would be proportional to the amount of particle production (namely, the sourced gauge fields in our model) generated during inflation. The LISA one instead would be sensitive to both the inflationary particle production, and the merging and accretion processes that occur between the PBH formation, and the present universe. The measurement of both signals could therefore allow us to probe the evolution between the PBH seeds and the present BH. 

If a stochastic GW background will be detected, the main challenge for claiming a cosmological origin will be to discriminate it against a possible astrophysics background. The GW signal sourced by this mechanism has two distinctive properties. One is its chirality, due to the preferential growth of one polarization w.r.t. the other one.  The prospect of detection of a chiral GW signal from a network of interferometers was studied in Refs.~\cite{Seto:2007tn,Crowder:2012ik,Smith:2016jqs}. The second is its nearly $\calO$ non-gaussianity \cite{Cook:2013xea,Shiraishi:2013kxa} (the bispectrum being about its power spectrum to the 3/2 power), which can also be probed by interferometers \cite{Thrane:2013kb}. 

To summarize, the GW signal from the pseudo-scalar interaction studied in this paper is a very natural candidate for the searches of a stochastic GW background on earth and space interferometers, due to the strong motivation of models of axion inflation, and the natural growth of the sourced signal at small scales. This potentially offers a window on scales of inflation on which we currently have little or no direct experimental knowledge. This signal has very characteristic properties (chirality, and order one non-gaussianity), which can help us discriminate it from an astrophysical background. The detectability of this signal requires that the PBH limit on scalar perturbations, that are unavoidably sourced together with the GW, are respected, possibly along the lines considered in this work.

\vskip.25cm
\noindent{\bf Acknowledgements:} 

We thank Yacine Ali-Ha\"imoud,  Jens Chluba, S\'ebastien Clesse, Andrei Linde,  Kin-Wang Ng, Lorenzo Sorbo, and Peter Tinyakov   for useful discussions.  We thank an anonymous referee for suggesting to include PTA experiments in our analysis. The work of M.P.  is partially supported from the DOE grant DE-SC0011842  at the University of Minnesota. The work of C.U. is supported by a Doctoral Dissertation Fellowship from the Graduate School of the University of Minnesota. The work of JGB is supported by the Research Project of the Spanish MINECO, FPA2013-47986-03-3P and FPA2015-68048-C3-3-P, and the Centro de Excelencia Severo Ochoa Program SEV-2012-0249.

\appendix

\section{Non-Gaussian scalar modes, and PBH formation }
\label{app:zetaPBH}

In this Appendix we discuss the necessary steps to obtain the upper limit in the right panel of Figure \ref{fig:limits} from the curve in the left panel. Namely, we discuss how the PBH mass is related to the number of e-folds when a mode left the horizon, and how the fraction $\beta$ is related to the primordial scalar power spectrum $P_\zeta$. We do this in two separate parts. 

\subsection{$M-N$ relation} 
\label{app:MN}

We derive here the relation between the number of e-folds $N$ before the end of inflation when a mode leaves the horizon, and the mass of the PBH that can be formed by this mode, if it has a large enough amplitude \cite{GarciaBellido:1996qt,Linde:2012bt}. We are interested in modes that re-enter the horizon during radiation domination. We assume that the radiation dominated era started right after inflation. We denote by $t_{\rm end}$ the end of inflation, and we normalize the scale factor to $a \left( t_{\rm end} \right) = 1$. 

Let us consider a density mode of physical wavelength $\lambda \left( t \right)$. We assume that this mode has a large enough amplitude to lead to a PBH when it re-enters the horizon after inflation. As customarily done, we take the inverse of the comoving momentum $k^{-1} = \frac{\lambda \left( t \right)}{2 \pi \, a \left( t \right)}$ of the mode as our best estimate for the comoving radius of the region associated to this mode that collapses to form the PBH. Therefore, the physical radius of this region at any given time is given by $R_k \left( t \right) = a \left( t \right) k^{-1}$.  

The comoving momentum of a mode  that exited the horizon $N$ e-folds before the end of inflation is 
\begin{equation}
k_N = a_N \, H_N = {\rm e}^{-N} \, H_N \;, 
\end{equation} 
where $a_N$ and $H_N$ are, respectively, the value of the scale factor and of the Hubble rate when the mode exits the horizon during inflation. 
Therefore 
\begin{equation}
R_{k_N} \left( t \right) = a \left( t \right) \,  {\rm e}^{N} \, H_N^{-1} \;.   
\label{Rk-t}
\end{equation}
The black hole mass is obtained from the  mass contained in this region when the mode re-enters the horizon, namely the mass in a sphere of radius $R_{k_N} \left( t = t_{\rm re-enter} \right)$.  

During radiation domination, $H = \frac{1}{2 t}$. Assuming radiation domination immediately from the end of inflation gives  $t_{\rm end} = \frac{1}{2 H_{\rm end,inf}}$, where $H_{\rm end,inf}$ is the Hubble rate at the end of inflation. Therefore, the scale factor during the radiation dominated era is given by 
\begin{equation}
a \left( t \right) = \left( \frac{t}{t_{\rm end}} \right)^{1/2} = \sqrt{2 \, H_{\rm end,inf} \, t} \;. 
\end{equation} 
The re-enter time is obtained by equating $H^{-1} \left( t_{\rm re-enter} \right)$ with $R_{k_N} \left( t = t_{\rm re-enter} \right)$. Using the above expressions we obtain $t_{\rm re-enter} = \frac{H_{\rm end,inf}}{2 \, H_N^2} \, {\rm e}^{2N} $. inserting this value in (\ref{Rk-t}), we find the physical radius we were looking for. Multiplying the volume of the corresponding sphere by the physical energy density at that time, $\rho \left( t_{\rm re-enter} \right) = 3 M_p^2 H^2  \left( t_{\rm re-enter} \right)$ we obtain the mass in that region. It is expected that a fraction $\gamma$ of this mass collapses into the black hole  \cite{Carr:2009jm}, giving the black hole mass 
\begin{equation}
M \simeq \gamma \,  4 \pi M_p^2 \, \frac{H_{\rm end,inf}}{H_N^2} \, {\rm e}^{2 N} \;, 
\end{equation}
or
\begin{equation}
\frac{M_{\rm BH}}{\rm g} = 13.3 \,  \gamma \, \frac{10^{13} \, {\rm GeV} \times H_{\rm end,inf}}{H_N^2} \, {\rm e}^{2 \, N} 
\label{MBH-N}
\end{equation}
The derivation we have just presented closely follows the analogous one in Ref.~\cite{Linde:2012bt}, that obtained  $\frac{M_{\rm BH}}{\rm g} \simeq 10  \, {\rm e}^{2 N}$. Compared with~\cite{Linde:2012bt}, we have also accounted for the variation of $H$ during inflation, and we have included the efficiency factor $\gamma$  \cite{Carr:2009jm}.~\footnote{We use the numerical value $\gamma = 3^{-3/2} \simeq 0.2$ suggested by the analytic computation of \cite{Carr:1975qj} for a collapse in the radiation dominated era (see \cite{Carr:2009jm} for a discussion).}

\subsection{$P_\zeta - \beta$ relation} 
\label{app:beta}

We derive here the relation between the curvature power spectrum $P_\zeta$, and the quantity $\beta$, which  is the fraction of regions collapsing to a PBH. A PBH is formed when a mode re-enters the horizon if the amplitude of this mode is above a certain threshold. Using the scalar curvature associated to this mode, the formation occurs if  $\zeta ( k_N ) \ga \zeta_{\rm c}$, where we recall that $k_N$ indicates the wavenumber corresponding to the mode that left the horizon N e-folds before the end of inflation. Therefore, the probability of forming a PBH is
\begin{equation}
\beta^{\rm form} \left( M_k \right) =   \int_{\zeta_c}^\infty {\cal P} \left( \zeta_k \right) \, d \zeta_k 
\label{betaformeq}
\end{equation} 
where ${\cal P} \left( \zeta_k \right)$ is a probability density for the scalar curvature $\zeta$. Since the primordial perturbations are Gaussian at CMB scales, it is common to assume that this probability is Gaussian. In the cases of interest in the present study, the scalar curvature is the sum of a vacuum part plus a part sourced by the gauge modes. The vacuum term is always negligibly small for PBH formation, and we are studying the formation due to the source term. This term originates from the convolution of two Gaussian modes, and it therefore obeys a $\chi^2$ statistics \cite{Linde:2012bt}. The PBH formation in the case of this distribution has been studied in Ref.~\cite{Lyth:2012yp}, and then also in \cite{Linde:2012bt,Byrnes:2012yx}. In this case, the expression (\ref{betaformeq}) gives (see for example Section IV of Ref.~\cite{Linde:2012bt} for details) 
\begin{equation}
\beta^{\rm form}_{\chi^2} (N)={\rm Erfc} \left( \sqrt{\frac{1}{2}+\frac{\zeta_c}{\sqrt{2P_{\zeta}(N)} } } \right) \;,  
\label{betaNG}
\end{equation}
where ${\rm Erfc} \left( x \right) \equiv 1 - {\rm Erf } \left( x \right)$ is the complementary error function. It is instructive \cite{Lyth:2012yp} to contrast this result with the case in which the perturbations are Gaussian (not our case): 
\begin{equation}
\beta^{\rm form}_{\rm Gaussian} (N)= \frac{1}{2} \, {\rm Erfc} \left(  \frac{\zeta_c}{\sqrt{2 P_{\zeta}(N)}} \right) \;\;. 
\label{betaG}
\end{equation}

\begin{figure}[tbp]
\centering 
\includegraphics[width=0.5\textwidth,angle=0]{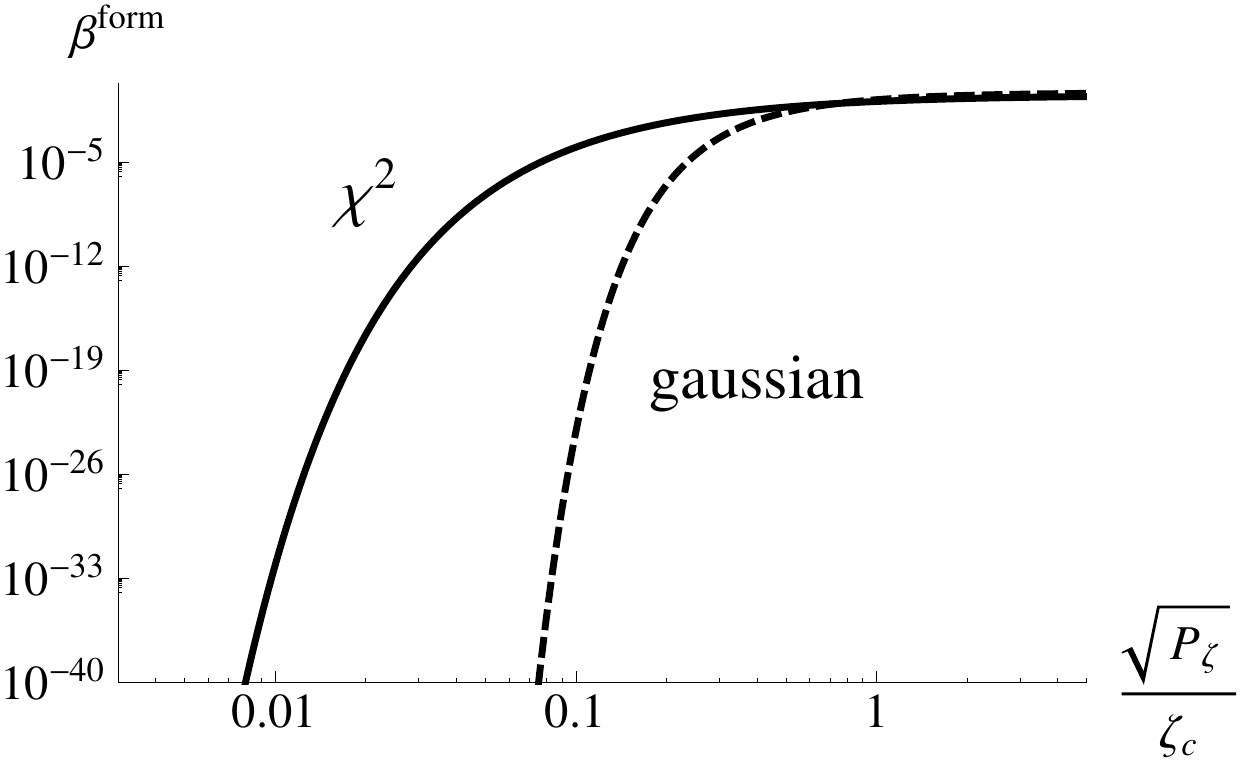}
\hfill
\caption{ 
Fraction of the universe that collapses to a PBH for any given mode, as a function of the value of the scalar curvature associated with that mode. 
In the regime of our interest, the same value of the power spectrum results in a much greater PBH fraction in the case of $\chi^2$ statistics (the one appropriate for the computations in this work) with respect to a Gaussian statistics.
}
\label{fig:Pz-beta}
\end{figure}

In Figure \ref{fig:Pz-beta} we show the fraction $\beta^{\rm form}$ as a function of the power spectrum, for both $\chi^2$ and Gaussian statistics of the primordial scalar perturbations. This result has been used to convert the PBH limits from the left to the right panel of Figure \ref{fig:limits}. Therefore, we are mostly interested in the values of $P_\zeta$ that result into a $\beta^{\rm form}$ in the $ \left[ 10^{-30} - 10^{-5} \right]$ interval. We see from the figure that, in this regime, a given value of the power spectrum results in a much bigger value of $\beta^{\rm form}$ in the $\chi^2$ vs. the Gaussian case. 

The result (\ref{betaNG}) provides the limit on the scalar perturbations from the PBH fraction $\beta$. In turn, this quantity is related to the parameter ${\tilde \beta}$ used in the left panel of Figure \ref{fig:limits} by the relation \cite{Carr:2009jm} 
\begin{equation}
{\tilde \beta} = \gamma^{1/2} \,\left( \frac{g_*}{106.75} \right)^{-1/4} \, \beta \;. 
\label{betatilde-beta}
\end{equation} 

In this expression, $g_*$ is the number of (effective bosonic) relativistic degrees of freedom in the thermal bath at the time in which a perturbation that gives rise to a PBH of mass $M$ re-enters the horizon, which we have normalized to the number of degrees of freedom in the Standard Model. From the relations written in Appendix \ref{app:MN}, the Hubble rate at the time of re-entry is related to the PBH mass $M$ by $H_{\rm re-enter} \simeq 4 \pi M_p^2 \gamma/M$. This corresponds to the temperature 
\begin{equation}
T_{\rm re-enter} \simeq 97 \, {\rm MeV} \, \left( \frac{106.75}{g_*} \right)^{1/4}  \sqrt{\frac{M_\odot}{M}} \,. 
\end{equation} 
We see from this relation that PBH masses $M \la 3 \cdot 10^{-7} \, M_\odot$ correspond to modes that re-enter at temperatures above the top quark mass, where the full Standard Model field content is relativistic. Using the relation (\ref{MBH-N}), with $H_N = H_{\rm end,inflation} = 10^{13} \, {\rm GeV}$ as a reference, this corresponds to $N \la 30$.  which is always the case for us, apart from the the discussion in Subsection~\ref{subsec:collision}. For this reason, when we convert the limits from ${\tilde \beta}$ to $P_\zeta$,  we simply fix $g_* = 106.75$ at all values of $N$. We note however that, strictly speaking,  the value of $g_*$ decreases for modes that exit the horizon at later times, corresponding to lower temperatures and smaller PBH masses. When the re-enter temperature drops well below the electron mass, one has $g_* = 3.36$. This happens for PBH masses $M \gg  2 \cdot 10^5 \, M_\odot$, corresponding to $N \gg 44$. In the worst case, our approximation introduces a mistake $\propto \left( \frac{106.75}{3.36} \right)^{1/4} \simeq 2.4$   in the value of $\beta$ (when used in Eq.~(\ref{betatilde-beta})), which propagates in a negligible way on the limit on $P_\zeta$  (we see from Figure \ref{fig:Pz-beta} that $\beta$ strongly depend on $P_\zeta$; therefore, inverting this dependence, $P_\zeta$ is very weakly dependent on $\beta$ in the regime of interest).

\section{Suppression factor ${\cal F}$} 
\label{app:calF}

In this Appendix we discuss and evaluate the factor ${\cal F}$ introduced in eq. (\ref{Pz2}). We start from the definition of the gauge invariant scalar curvature, evaluated in spatially flat gauge, $\zeta \equiv  - \frac{H \delta \rho}{\dot{\rho}}$. From the background equations of the model (\ref{lagrangian-phi}) one finds \cite{Notari:2016npn} $\dot{\rho} = - 3 H \dot{\phi}^2 - 4 H \rho_R$, where $\rho$ is the total energy density in the model, and  $\rho_R$ is the energy density in the vector field  \cite{Anber:2009ua} 
\begin{equation}
\rho_R = \frac{\langle \vec{E}^2 + \vec{B}^2 \rangle}{2} \simeq  1.4 \cdot 10^{-4} \frac{H^4}{\xi^3} {\rm e}^{2 \pi \xi} \;. 
\end{equation} 
We then find 
\begin{eqnarray}
\zeta =  - \frac{H \, \delta \phi}{\dot{\phi}} \times {\cal F} \;\;\;\;\;\;,\;\;\;\;\;\; 
{\cal F} = \frac{-\dot{\phi} \, V'}{H \left( 3 \dot{\phi}^2 + 4 \frac{\langle E^2+B^2 \rangle}{2} \right)} \;. 
\label{calF}
\end{eqnarray}

We have written the first equation as the  standard relation $\zeta =  - \frac{H \, \delta \phi}{\dot{\phi}} $, times a correction factor. The standard relation applies in the regime of negligible backreaction of the vector  field on the background evolution of the inflaton and of the scale factor, namely when $ \frac{\langle E^2+B^2 \rangle}{2} $ is negligible, and $3 H \dot{\phi} \simeq - V'$ (in which case, ${\cal F} = 1$). 
However, more in general, this factor needs to be included  \cite{Notari:2016npn}, and properly evaluated. Refs.  \cite{Anber:2009ua,Barnaby:2011qe,Linde:2012bt} did not include this effect. So, the power spectrum expression that we have given in (\ref{Pz2}) is the one considered in those works,  times a ${\cal F}^2$ correction. 
 
Ref. \cite{Notari:2016npn} argues that this corrections is very important when $\rho_R$ is much greater than the kinetic energy of the inflation field. This statement needs to be qualified. We note that 
\begin{equation}
{\cal F} \neq  \frac{3 \, \dot{\phi}^2}{3 \dot{\phi}^2 + 4 \frac{\langle E^2+B^2 \rangle}{2}} \;. 
\label{calF-wrong}
\end{equation} 
This relation is correct only in the regime of negligible backreaction, when ${\cal F} \rightarrow 1$. Ref.  \cite{Notari:2016npn} did not write such a relation. However, their argument that ${\cal F}$ is important when the energy density in the gauge field  is greater than the inflaton kinetic energy would be  guaranteed to be valid only if such a relation was  correct. In general, we need to evaluate ${\cal F}$ using the correct expression (\ref{calF}).

\begin{figure}
\centerline{
\includegraphics[width=0.35\textwidth,angle=0]{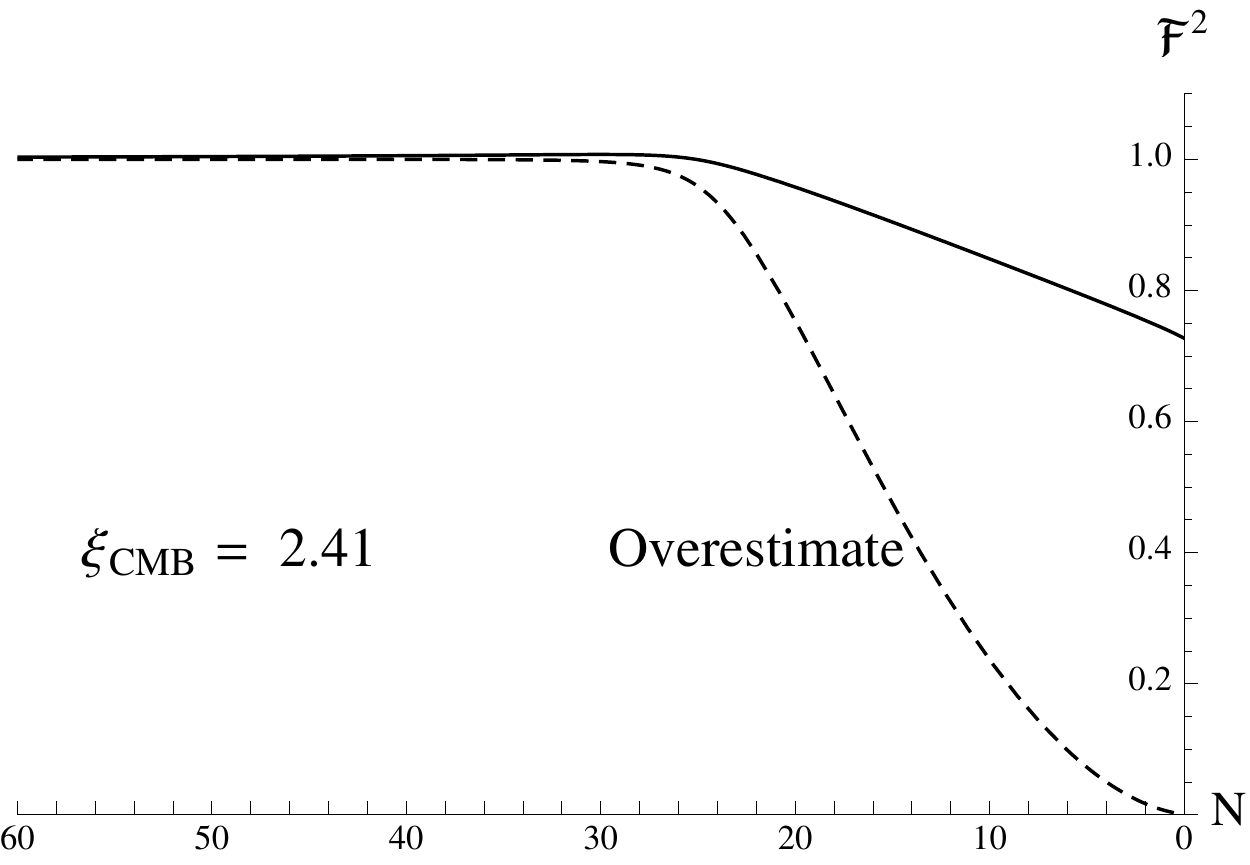}
}
\caption{Correction factor on the power spectrum as a function of the number of e-folds of inflation, for a linear potential, and for a coupling inflaton-gauge field that saturates the Planck bound. The solid line is obtained from eq. (\ref{calF}) and it is the correct way to evaluate the correction. The dashed line uses the incorrect expression (\ref{calF-wrong}). 
}
\label{fig:Fcal2}
\end{figure}

We do so in Figure \ref{fig:Fcal2} where, as in the main text, we have chosen a linear inflaton potential, and a coupling to the gauge field as large as that allowed by the CMB data in the model that we are considering. The solid line shown in the figure is the quantity ${\cal F}^2$ evaluated through the proper expression (\ref{calF}). On the other hand, the dashed line is obtained from the incorrect evaluation of ${\cal F}$ using eq. (\ref{calF-wrong}), which makes use of the ratio between the energy in the vector field and the kinetic energy of the inflaton. We see that this second case results in a much greater departure of ${\cal F}$ from one, and therefore in a great overestimate of the effect of the correction factor on the scalar power spectrum. 

In fact, we can obtain the asymptotic value of ${\cal F}$ analytically in the regime of strong backreaction.~\footnote{We thank Lorenzo Sorbo for discussions on this issue, and for suggesting this analytical computation.}  In this case, one has  \cite{Anber:2009ua} 
\begin{equation}
\frac{ \langle \vec{E}^2 + \vec{B}^2 \rangle}{2} \simeq - \frac{4}{7} \, \xi \, \langle \vec{E} \cdot \vec{B} \rangle \simeq - \frac{4}{7} \, \xi \, f \, V' \;. 
\end{equation} 
Inserting this in (\ref{calF}), and disregarding the $3 \dot{\phi}^2$ contribution to the denominator (which is appropriate in the regime of strong backreaction), we immediately obtain ${\cal F} \simeq \frac{7}{8}$, in excellent agreement with the late time behavior of the solid line of Figure 
\ref{fig:Fcal2}. This is the asymptotic value that is reached in the limit of strong backreaction, independently of the inflaton potential, of the coupling $f$, and of the number of gauge fields amplified by this mechanism.


\end{document}